\documentclass[10pt,reqno,fleqn]{amsart}
\usepackage{graphicx}
\newtheorem{thm}{Theorem}[section]

\newtheorem{prop}[thm]{Proposition}
\theoremstyle{definition}

\theoremstyle{remark}

\numberwithin{equation}{section}

\newcommand{\M}{\mathcal{M}}
\newcommand{\Z}{\mathcal{Z}}

\newcommand{\J}{\mathcal{J}}

\newcommand{\F}{\mathcal{F}}
\begin{document}
\title[
R-H approach for correl. of characteristic polynomials ]{}
\maketitle
\begin{flushleft}
\begin{huge}
{\textbf{Universal Results for Correlations of\\[10pt]
Characteristic Polynomials:\\[10pt]
Riemann-Hilbert Approach}}
\end{huge}\\[12pt]
\end{flushleft}
{\textbf{Eugene Strahov,  Yan V. Fyodorov}}\\[10pt]
Department of Mathematical Sciences, Brunel University,
Uxbridge,\\
UB8 3PH, United Kingdom\\
e-mail: Eugene.Strahov@brunel.ac.uk, Yan.Fyodorov@brunel.ac.uk\\[20pt]
{\textbf{Abstract}} We prove that general correlation functions of
both ratios and products of characteristic polynomials of
Hermitian random matrices are governed by integrable kernels of
three different types: a) those constructed from orthogonal
polynomials; b) constructed from Cauchy transforms of the same
orthogonal polynomials and finally c) those constructed from both
orthogonal polynomials and their Cauchy transforms. These kernels
are related with the Riemann-Hilbert problem for orthogonal
polynomials. For the correlation functions we obtain exact
expressions in the form of determinants of these kernels. Derived
representations enable us to study asymptotics of correlation
functions of characteristic polynomials via Deift-Zhou
steepest-descent/stationary phase method for Riemann-Hilbert
problems, and in particular to find negative moments of
characteristic polynomials. This reveals the universal parts of
the correlation functions and moments of characteristic
polynomials for arbitrary invariant ensemble of $\beta=2$ symmetry
class.
\\[20pt]
\begin{flushleft}
\section{\textbf{Introduction}}
\end{flushleft} Correlation functions of characteristic polynomials
for various  ensembles of random matrices were investigated by a
number of authors in a series of recent papers. Keating and Snaith
\cite{KS}, Hughes, Keating and O'Connell \cite{keating1,keating11}
demonstrated that averages of characteristic polynomials over
ensembles of random matrices can be useful to make predictions
about moments of Riemann zeta function, and other L-functions.
These authors consider ensembles of matrices associated with
compact groups (the simplest case in the family of ensembles of
$\beta=2$ symmetry class) and derive moments of characteristic
polynomials \cite{KS}. In subsequent papers (Conrey, Farmer,
Keating, Rubinstein and Snaith \cite{keating3,keating4}) compute
more general (autocorrelations or "shifted moments") correlation
functions of \textit{products} of characteristic polynomials.

Brezin and Hikami \cite{brezin1} (and also Mehta and Normand
\cite{MN}) considered correlation functions of products of
characteristic polynomials for an arbitrary unitary invariant
ensemble of \textit{Hermitian} matrices. This family of ensembles
is characterized by the weight $\exp\left[-N\mbox{Tr}V(H)\right]$
in the corresponding probability measure ( $V(H)$ is  an
essentially arbitrary potential function, $N$ is the dimension of
the matrix $H$). Using the method of orthogonal polynomials, they
found both exact and asymptotic (large $N$) expressions for the
correlation functions, and for the positive moments.
This enabled them to investigate
universality of results and find dependence on density of states.
Namely, the asymptotic expressions were proved to be factorized in
product of  universal and non-universal (ensemble-dependent)
parts. It was found that the universal numerical pre-factors of
positive moments of characteristic polynomials coincide
asymptotically with those  for the unitary random matrices
obtained by Keating and Snaith \cite{KS}. Thus it was rather
naturally to expect that these universal pre-factors should appear
in the positive moments of Riemann zeta function.

While Brezin and Hikami compare the positive moments of
characteristic polynomials with the positive moments of Riemann
zeta function, it is similarly worth to compare negative moments
of characteristic polynomials with the negative moments of zeta
function. Note that such comparison makes sense only if the degree
of universality of the negative moments is established. Indeed, it
is clear that only universal parts of moments of characteristic
polynomials (universal pre-factors, for example) may be related
with the corresponding moments of zeta function.

 For the negative moments of zeta function a conjecture is
available due to
 the work by Gonek \cite{gonek}. Fyodorov in \cite{I} performed
 the calculations for negative moments of characteristic
 polynomials for
the simplest case of Hermitian random matrices with the Gaussian
potential function $V(H)=H^2/2$, known as the Gaussian Unitary
ensemble (GUE). The result in \cite{I} agreed with those by Gonek
\cite{gonek}. However, a full comparison was still not possible
since  the universal results were unavailable.

In this paper we both find the negative moments of characteristic
polynomials for any unitary invariant ensemble of Hermitian
matrices and compare them with moments of zeta function.

A study of the negative moments of characteristic polynomials is
further motivated by the recent observation by Berry and Keating
in \cite{berry}. These authors argue that divergences of the
negative moments could be determined by degeneracies in the
spectrum, or clusters of eigenvalues. It is an interesting
assertion as clusters should be rare events for a random matrix
due to the level repulsion. Berry and Keating show that the
question of whether the influence of clusters is dominant is
related to that how the negative moments diverge.  More precisely,
while the negative moments of characteristic polynomials are
divergent this divergence can be removed once we agree  that we
consider these moments off the real axis (i.e. on the line shifted
from the real axis by small parameter $\delta$). Then the moments
are well defined and, in principle, can be computed. According to
Berry and Keating scenario  it follows that the $2K$ negative
moments are proportional to $\delta^{-K^2}$ as $\delta$ goes to
zero for unitary invariant ensembles of Hermitian matrices.
 In \cite{I} it was shown that indeed the
negative moments diverge as $\delta^{-K^2}$  for the case of GUE.
In this paper we prove the universality of this result, i.e.
negative moments of characteristic polynomials diverge as
$\delta^{-K^2}$ for all unitary invariant ensembles of Hermitian
matrices.

Another important class of correlation functions includes (product
of) \emph{ratios} of characteristic polynomials. As is well known
those correlation functions can be used to extract more
conventional
 n-point correlators of the spectral densities (see,
for example, \cite{Haake,chir} and the references therein).  As a
simple illustration of this statement we compute correlation
functions which include \emph{ratios} of characteristic
polynomials for an arbitrary unitary invariant ensemble, and then
reproduce a well known asymptotic result for the two-point
correlation function of the resolvents.

An even more general class of correlation functions are those that
include both products and ratios of characteristic polynomials
(i.e. when the numbers of characteristic polynomials in the
numerator is different from that in the denominator). These
functions provide a very detailed information about spectra of
random matrices. For this reason such correlation functions
(together with autocorrelation functions of characteristic
polynomials) are pervasively used in the field of Quantum Chaos
see \cite{AS,Efetov,Haake,II} and references therein. For example,
they are used for extracting generating functions for such
physically interesting characteristics as the distributions of the
"local" density of states and of the "level curvatures"  (see e.g.
Andreev and Simons \cite{AS} for more detail).

For the particular case of the Gaussian Unitary Ensemble (GUE) the
large $N$ asymptotics of such  correlation functions  is known. A
straightforward way to study the asymptotics is to exploit the
supersymmetry technique and its modifications
\cite{AS,szabo,I,II}. For example, in \cite{I} one can find
asymptotic values for negative moments for the GUE, and Andreev
and Simons were the first who obtained the asymptotics for other
correlation functions. Moreover, Brezin and Hikami in their recent
paper \cite{brezin3} have made an effort to apply the supersymmetric
technique to correlations of ratios of characteristic polynomials
for more complicated Gaussian Orthogonal and Gaussian Symplectic
ensembles.

However, the rigorous application of the supersymmetry technique
(and its various modifications) is limited to Gaussian ensembles only.
Another common disadvantage of these approaches is that they
 are rather robust. While they work well for
 the computation of asymptotics, investigation of
the correlation functions for finite size matrices
is hardly possible.  Moreover, those methods
hid a nice determinantal structure (revealed in our recent paper
\cite{strahov1}) of the exact expressions and only yielded the
asymptotic result in a form of a sum over permutations
\cite{AS,II}.

Thus, for understanding the correlation functions of more
general (non-Gaussian) ensembles a different procedure is
required. In the present paper we solve the above mentioned problems and
provide a unified approach to general correlation functions of
characteristic polynomials for unitary invariant ensembles of
Hermitian matrices.
\begin{flushleft}\end{flushleft}
\section{\textbf{Statement of the Problem and the Main Results}}
\begin{flushleft}\end{flushleft}
\nopagebreak[3] Let $H$ denote a $N\times N$ random Hermitian
matrix which is an element of a unitary invariant ensemble (i.e.
that of $\beta=2$ symmetry class). Introducing $N$-dimensional
vector $\hat x$ of eigenvalues of the matrix $H$, one defines the
ensemble by the eigenvalue density function $P_{N}(\hat x)$ (see
Mehta \cite{mehta}),
\begin{equation}\label{EigenvalueDensity}
P_{N}(\hat
x)=\left[Z_N\right]^{-1}\exp\left[-N\sum\limits_{1}^NV(x_i)\right]\triangle^2(\hat
x)
\end{equation}
   The symbol $\triangle(\hat x)$ stands for the
Vandermonde determinant, $V(x)$ is a potential function and $Z_N$
is a normalization constant.

A \textit{characteristic polynomial}
which corresponds to the matrix $H$ is defined as
$\Z_{N}\left[\epsilon,H\right]=\mbox{det}\left(\epsilon-H\right)$.
This object is a building block for constructing various
correlation functions of interest, such as correlation functions
of products and ratios of characteristic polynomials. Our main goal
 is to provide a systematic method for
computing all these correlation functions for
non-Gaussian ensembles, i.e. for the potential $V(x)$ which is more
general then $x^2$.

From this end we first consider the correlation functions of the following
types:
\begin{equation}\label{CorrelationOfProducts}
\F_I^K(\hat\lambda,\hat\mu)=\left\langle
\prod\limits_{1}^{K}\Z_N\left[\lambda_j,H\right] \Z_N\left[\mu_j,
H\right] \right\rangle_{H}\,\,,
\end{equation}
\begin{equation}\label{CorrelationOfRatios}
\F_{II}^K(\hat\mu, \hat\epsilon)=\left\langle \prod\limits_{1}^{K}
\frac{\Z_N\left[\mu_j,H\right]}{\Z_N\left[\epsilon_j,H\right]}
\right\rangle_{H}\,\,,
\end{equation}
\begin{equation}\label{CorrelationOfNegative}
\F_{III}^K(\hat\epsilon,\hat\omega)=\left\langle
\prod\limits_{1}^{K}\frac{1}{\Z_N\left[\epsilon_j,
H\right]\Z_N\left[\omega_j, H\right]}\right\rangle_{H}\,\,,
\end{equation}
where the averages are understood as integrals with respect to the
measure $d\mu(\hat x)=P_N(\hat x)d\hat x$, with $P_N(\hat x)$ being
defined by the equation (\ref{EigenvalueDensity}). If the components
of vectors $\hat\epsilon$ and $\hat\omega$ have nonzero imaginary
parts the correlation functions above are well defined.

Our first
result is that each of the above correlation functions are
essentially governed
by two-point kernels constructed from
monic orthogonal polynomials $\pi_k(x)$ and their Cauchy
transforms, $h_k(\epsilon)$. The monic polynomials,
$\pi_j(x)=x^j+\ldots$, orthogonal
 with respect to
the measure $d\mu(x)=e^{-NV(x)}dx$, are defined by
\begin{equation}
\int\pi_k(x)\pi_m(x)e^{-NV(x)}dx=c_kc_m\delta_{km}
\end{equation}
and their Cauchy transforms are determined in accordance with the
following expression
\begin{equation}\label{CauchyTransform}
h_k(\epsilon)=\frac{1}{2\pi i}\int\frac{e^{-
NV(x)}\pi_k(x)dx}{x-\epsilon}, \;\;\;\epsilon\in
\mathbb{C}/\mathbb{R}
\end{equation}
The correspondence between the types of correlation functions and
different kernels obtained in this paper is summarized in Table 1.
\begin{table}[h]
\begin{tabular}{|c|c|}
 \hline
 & \\
  Correlation function & Kernel \\
  & \\
  \hline
   & \\
$\F_I^K(\hat\lambda,\hat\mu)$ &  $
  W_{I,N+K}(\lambda,\mu)=
  \frac{\pi_{N+K}(\lambda)\pi_{N+K-1}(\mu)-\pi_{N+K-1}(\lambda)\pi_{N+K}(\mu)
  }{\lambda-\mu}\nonumber $ \\
  & \\
  \hline
   & \\
  $\F_{II}^K(\hat\epsilon,\hat\mu)$ & $
  W_{II,N}(\epsilon,\mu)=
  \frac{h_{N}(\epsilon)\pi_{N-1}(\mu)-h_{N-1}(\epsilon)\pi_{N}(\mu)
  }{\epsilon-\mu}\nonumber $ \\
 & \\
   \hline
   & \\
  $\F_{III}^K(\hat\epsilon,\hat\omega)$ & $
  W_{III,N-K}(\epsilon,\omega)=
  \frac{h_{N-K}(\epsilon)h_{N-K-1}(\omega)-h_{N-K-1}(\epsilon)h_{N-K}(\omega)
  }{\epsilon-\omega}\nonumber $ \\
 & \\
   \hline
\end{tabular}
\small{\caption{Correlation functions and kernels}}
  \end{table}\\
One of those kernels, the kernel $W_{I,N+K}(\lambda,\mu)$, is well
known in the theory of random
matrices. It is related to the familiar kernel $K_{N}(\lambda,\mu)$
which is known to determine completely the n-point correlation
functions of eigenvalue densities as well as
spacing distributions between eigenvalues.
The kernel $K_{N}(\lambda,\mu)$ is defined by
\begin{equation}\label{standardkernel}
K_N(x,y)= -\frac{\gamma_{N-1}}{2\pi
i}\;e^{-\frac{N}{2}V(x)}\frac{\pi_N(x)\pi_{N-1}(y)-
\pi_{N-1}(x)\pi_N(y)}{x-y}\;e^{-\frac{N}{2}V(y)}
\end{equation}
where
\begin{equation}\label{gamma}
\gamma_{n-1}=-\frac{2\pi i}{c^2_{n-1}}
\end{equation}
However two other kernels $W_{II,N}(\epsilon,\mu)$,
$W_{III,N-K}(\epsilon,\omega)$ responsible for the correlation
functions of characteristic polynomials have not been previously
considered, to the best of our knowledge.

Apart from the correlation functions discussed above
  we investigate also the correlation functions containing non-equal
  number of characteristic polynomials in the numerator and
the denominator, such as
\begin{equation}\label{CorrelationOfRatiosL}
\F_{IV}^{K,M}(\hat\epsilon, \hat\mu)=\left\langle
\frac{\prod\limits_{1}^{K}
\Z_N\left[\mu_l,H\right]}{\prod\limits_{1}^{M}\Z_N\left[\epsilon_j,H\right]}
\right\rangle_{H},\;\;\;0<M<K
\end{equation}
and
\begin{equation}\label{CorrelationOfRatiosS}
\F_{V}^{K,M}(\hat\epsilon, \hat\mu)=\left\langle
\frac{\prod\limits_{1}^{K}
\Z_N\left[\mu_l,H\right]}{\prod\limits_{1}^{M}\Z_N\left[\epsilon_j,H\right]}
\right\rangle_{H},\;\;\;0<K<M
\end{equation}
where $K+M$ is an even number, i.e. $K+M=2L$. We reveal that these
functions can be expressed in terms of determinants of size $L\times L$. The
entries of the determinant for the function
$\F_{IV}^{K,M}(\hat\epsilon, \hat\mu)$ are kernels $W_{II,N-M+L}$
and $W_{I,N-M+L}$ while the entries of the corresponding
determinant for the function $\F_{V}^{K,M}(\hat\epsilon, \hat\mu)$
are the kernels $W_{II,N-M+L}$ and $W_{III,N-M+L}$.

It is important to note that all the kernels
introduced above belong to the family of
the so-called "integrable" kernels,
i.e. they correspond to  integrable operators first distinguished
as a class by Its, Izergin , Korepin and Slavnov
\cite{itsizergin1,itsizergin2,itsizergin3}.
According to that theory the kernels corresponding to the
integrable operators are defined as follows: Let $\Sigma$ be oriented contour in
$\mathbb{C}$. An operator $L$ acting in $L^2(\Sigma, \mathbb{C})$
is called integrable if its kernel has the form
\begin{equation}\label{IntegrableForm}
L(z,z')=\frac{\sum\limits_{j=1}^{M}f_j(z)g_j(z')}{z-z'}
\end{equation}
for some functions $f_i, g_j ;\;\; i,j=1,\ldots , M$. The
formalism of integrable operators is described by Deift in
\cite{deift0}.

Our kernels obviously satisfy the above definition.
This is an important observation as integrable kernels have a
variety of useful properties that can be exploited (see a review
paper by Deift, Zhou and Its \cite{univ1}). For example, it is
known that the Fredholm determinants of integrable kernels satisfy
non-linear differential equations. In particular, Tracy and
Widom \cite{tracy1}-\cite{tracy4} obtained differential equations
(of the Painlev\'{e} type) for the Fredholm determinants
associated with the kernel $K_N(x,y)$. In the present paper we use
another important property of integrable kernels, namely, their
relation with Riemann-Hilbert problems. In a similar way as the
Riemann-Hilbert technique was applied to the kernel $K_N(x,y)$ to
demonstrate its universality in the Dyson's scaling limit (see
\cite{deift1}, \cite{univ1} and \cite{bleher})) we exploit the
Riemann-Hilbert approach to find Dyson's scaling limit of kernel
functions $W_{I,N+K}(\lambda,\mu)$, $W_{II,N}(\epsilon,\mu)$ and
$W_{III,N-K}(\epsilon,\omega)$. As a result we obtain associated
\emph{universal} kernels summarized in Table 2.
\begin{table}[h]
\begin{tabular}{|c|c|}
 \hline
 & \\
 Finite $N$ kernel functions & Associated limiting kernels\\
  & \\
  \hline
   & \\
$W_{I,N+K}(\lambda,\mu)$ &  $
  {\mathbb{S}}_I(\zeta-\eta)=\frac{\sin\left[\pi(\zeta-
  \eta)\right]}{\pi(\zeta-\eta)}$ \\
  & \\
  \hline
   & \\
  $W_{II,N}(\epsilon,\mu)$, & $
  {\mathbb{S}}_{II}(\zeta-\eta)=\begin{cases}
    \frac{e^{i\pi(\zeta-\eta)}}{\zeta-\eta}, & \Im m\;\zeta>0 \\
    \frac{e^{-i\pi(\zeta-\eta)}}{\zeta-\eta} & \Im m\;\zeta<0
  \end{cases}
  $
  \\
   & \\
$\Im m\;\epsilon\neq 0$ & \\
   \hline
   & \\
  $W_{III,N-K}(\epsilon,\omega)$ & ${\mathbb{S}}_{III}(\zeta-\eta)=\begin{cases}
    \frac{1}{\zeta-\eta}, & \Im m\;\zeta>0,\; \Im m\;\eta<0\\
    \frac{-1}{\zeta-\eta} & \Im m\;\zeta<0,\; \Im m\;\eta>0\\
    0 & \mbox{otherwise}
  \end{cases}
  $ \\
 & \\
 $\Im m\;\epsilon\neq 0,\;\Im
 m\;\omega\neq 0$ &  \\
   \hline
\end{tabular}
\\
\small{\caption{Finite and associated limiting kernels}}
  \end{table}
\begin{flushleft}\end{flushleft}
The representation of the correlation functions
in terms of determinants of the kernels (see
section 4)  enables us to give explicit asymptotic formulae for all five
correlation functions of characteristic polynomials discussed in the text
above. We give a summary of these results below.
\begin{flushleft}\end{flushleft}
\emph{2.1 Dyson's Limit for
$\F_{I}^K(\hat\lambda,\hat\mu)=\left\langle\prod\limits_{1}^{K}\Z_N\left[\lambda_j,H\right]
\Z_N\left[\mu_j, H\right] \right\rangle_{H}$}
\begin{flushleft}\end{flushleft}
\nopagebreak[3]Define $K$-dimensional vectors,
  $\hat x=(x,\ldots ,x)$, where $x$ belongs to the support of the equilibrium measure
  for the potential function $V(x)$ (see section 5 for the definitions), $\hat\zeta=(\zeta_1,\ldots,\zeta_K)$ and
$\hat\eta=(\eta_1,\ldots,\eta_K)$. Then for the
  correlation function of products of characteristic polynomials
    we obtain
\begin{equation}
\F_{I}^K(\hat
  x+\hat\zeta/N\rho(x),\hat x+\hat\eta/N\rho(x))\nonumber
\end{equation}

\begin{equation}\label{AsymptoticFI}
\qquad=\left[c_N\right]^{2K}\;\frac{e^{KNV(x)}\left[N\rho(x)\right]^{K^2}e^{\alpha(x)\sum\limits_1^K(\zeta_l+\eta_l)}}
{\triangle(\hat\zeta)\triangle(\hat\eta)}\;\mbox{det}\left[
\mathbb{S}_{I}(\zeta_i-\eta_j)\right]_{1\leq i,j\leq K}
\end{equation}
where $\rho(x)$ stands for the density of states, and we have
introduced the notation
\begin{equation}\label{ALPHAX}
\alpha(x)=\frac{V'(x)}{2\rho(x)}
\end{equation}
\begin{flushleft}\end{flushleft}
\emph{2.2 Dyson's Limit for $\F_{II}^K(\hat\epsilon,
\hat\mu)=\left\langle \prod\limits_{1}^{K}
\frac{\Z_N\left[\mu_j,H\right]}{\Z_N\left[\epsilon_j,H\right]}
\right\rangle_{H}$}
\begin{flushleft}\end{flushleft}
 Let $\hat x, \hat\zeta, \hat\eta$ be $K$ -dimensional vectors.
 Assume that the  components of $\hat\zeta$ have non-zero
 imaginary parts. Then we find
 \begin{equation}\label{DysonLimitFII}
\F_{II}^K(\hat
  x+\hat\zeta/N\rho(x),\hat x+\hat\eta/N\rho(x))
  \end{equation}\nopagebreak[3]
\begin{equation}
 \qquad\qquad\qquad =
(-)^{\frac{K(K-1)}{2}}\;
e^{-\alpha(x)\sum\limits_1^K(\zeta_l-\eta_l)}
\frac{\triangle(\hat\zeta,\hat\eta)}{\triangle^2(\hat\zeta)
\triangle^2(\hat\eta)}\;\mbox{det}
\left([\mathbb{S}_{II}(\zeta_i-\eta_j)\right]_{1\leq i,j\leq
K}\nonumber
\end{equation}
\begin{flushleft}\end{flushleft}
\emph{2.3 Dyson's limit of $\F_{III}^K(\hat\varpi,
\hat\omega)=\left\langle \prod\limits_{1}^{K}
\frac{1}{\Z_N\left[\varpi_j,H\right]\Z_N\left[\omega_j,H\right]}
\right\rangle_{H}$}
\begin{flushleft}\end{flushleft}\nopagebreak[3]
It is convenient to introduce $2K$ dimensional vector,
$\hat\epsilon=(\hat\varpi,\hat\omega)$.  The new coordinates
appropriate for investigation of Dyson's asymptotic limit of the
correlation function are defined so that $\hat\epsilon=\hat
x+\hat\zeta/N\rho(x)$, $\dim \hat x=\dim \hat\zeta=2K$. Here the
vector $\hat x$ has $2K$ equal components $x$, and $x$ belongs
to the support of the equilibrium measure for the potential
function $V(x)$. As for the components of the vector $\hat\zeta$, we
assume that they have non-zero  imaginary parts. We find
\begin{equation}
\F_{III}^K(\hat x+\hat\zeta/N\rho(x))=(-)^K\left[\gamma_N\right]^K
\left[N\rho(x)\right]^{K^2}e^{-KNV(x)}e^{-\alpha(x)\sum\limits_1^K(\zeta_l+\zeta_{l+K})}
\nonumber
\end{equation}

\begin{equation}\label{ASYMPTOTICFIII}
\qquad\times\frac{1}{(2K)!}\;\sum\limits_{\pi\;\in\;
{\textsf{S}}_{2K}}
\frac{\mbox{det}\left[\mathbb{S}_{III}\left(\zeta_{\pi(i)}-
\zeta_{\pi(j+K)}\right)\right]_{ 1\leq i,j\leq
K}}{\triangle(\zeta_{\pi(1)},\ldots,\zeta_{\pi(K)})
\triangle(\zeta_{\pi(K+1)}, \ldots,\zeta_{\pi(2K)})}
\end{equation}
\begin{flushleft}\end{flushleft}
\emph{2.4 Dyson's limit for $\F_{IV}^K(\hat\epsilon,
\hat\mu)=\left\langle \frac{\prod\limits_{1}^{K}
\Z_N\left[\mu_l,H\right]}{\prod\limits_1^M\Z_N\left[\epsilon_j,H\right]}
\right\rangle_{H},\;\;K>M $}
\begin{flushleft}\end{flushleft}\nopagebreak[3]
In this case  new coordinates are introduced so that
$\hat\epsilon=\hat x+\hat\zeta/N\rho(x)$, $\hat\mu=\hat
x+\hat\eta/N\rho(x)$. It is clear that $\mbox{dim}\; \hat\zeta
=M$, $\mbox{dim}\; \hat\eta=K$, $\Im m\;\zeta\neq 0$.  We define
$2L=K+M$ (i.e. we consider the correlation function of an even
number of characteristic polynomials). With these definitions we
find
\begin{equation}
\F_{IV}^K(\hat x+\hat\zeta/N\rho(x),\hat
x+\hat\eta/N\rho(x))\nonumber
\end{equation}

\begin{equation}
\qquad\qquad\quad\;=(-)^{\frac{M(K-1)}{2}}\left[c_N\right]^{K-M}\left[N\rho(x)\right]^{(L-M)^2}
e^{N(L-M)V(x)}
\end{equation}

\begin{equation}
\qquad\qquad\quad\;\times\;
e^{-\alpha(x)\left[\sum\limits_1^M\zeta_l-\sum\limits_1^K\eta_l\right]}
\frac{\triangle(\zeta_1, \ldots,\zeta_M;\eta_{L-M+1},
\ldots,\eta_K)}{\triangle^2(\hat\zeta)\triangle^2(\eta_{L-M+1},\ldots,\eta_K)
\triangle(\eta_1,\ldots ,\eta_{L-M})}\nonumber
\end{equation}

\begin{equation}
\qquad\qquad\;\;\;\;\;\times\;\mbox{det} \left|
\begin{array}{ccc}
  \mathbb{S}_{II}(\zeta_1-\eta_{L-M+1}) & \ldots & \mathbb{S}_{II}(\zeta_1-\eta_K) \\
  \vdots &  & \\
  \mathbb{S}_{II}(\zeta_M-\eta_{L-M+1}) & \ldots & \mathbb{S}_{II}(\zeta_M-\eta_K) \\
\mathbb{S}_{I}(\eta_1-\eta_{L-M+1}) & \ldots & \mathbb{S}_{I}(\eta_1-\eta_K) \\
  \vdots &  & \\
 \mathbb{ S}_{I}(\eta_{L-M}-\eta_{L-M+1}) & \ldots & \mathbb{S}_{I}(\eta_{L-M}-\eta_K) \\
\end{array}
\right| \nonumber
\end{equation}
\begin{flushleft}\end{flushleft}
\emph{2.5 Dyson's limit for $\F_{V}^K(\hat\epsilon,
\hat\mu)=\left\langle \frac{\prod\limits_{1}^{K}
\Z_N\left[\mu_l,H\right]}{\prod\limits_1^M\Z_N\left[\epsilon_j,H\right]}
\right\rangle_{H},\;\;K<M $}
\begin{flushleft}\end{flushleft}\nopagebreak[3]
We introduce  new coordinates $\hat x$, $\hat \zeta$, $\hat\eta$
as in the previous case. Let $2L=K+M$. Then the Dyson's limit of
the correlation function $\F_{V}^K(\hat x+\hat\zeta/N\rho(x),\hat
x+\hat\eta/N\rho(x))$ is
\begin{equation}
\F_{V}^K(\hat x+\hat\zeta/N\rho(x),\hat
x+\hat\eta/N\rho(x))\nonumber
\end{equation}

\begin{equation}\label{asymptoticFV}
=(-)^{\frac{M(M-1)}{2}}\left[\gamma_N\right]^{M-L}\left[N\rho(x)\right]^{(L-K)^2}
e^{N(L-K)V(x)}\frac{1}{\triangle^2(\hat\eta)} \frac{1}{M!}
\end{equation}

\begin{equation}
\times\;e^{-\alpha(x)\left[\sum\limits_1^M\zeta_l-\sum\limits_1^K\eta_l\right]}
\sum\limits_{\pi\;\in\; {\textsf{S}}_M}
 \frac{\triangle(\eta_1, \ldots,\eta_K;\zeta_{\pi(1+\frac{M-K}{2})},
\ldots,\zeta_{\pi(M)})}{\triangle^2(\zeta_{\pi(1+\frac{M-K}{2})},
\ldots,\zeta_{\pi(M)}) \triangle(\zeta_{\pi(1)},
\ldots,\zeta_{\pi(\frac{M-K}{2})})}\nonumber
\end{equation}

\begin{equation}
\times\;\mbox{det} \left|
\begin{array}{ccc}
  \mathbb{S}_{II}(\zeta_{\pi(1+\frac{M-K}{2})}-\eta_{1}) & \ldots & \mathbb{S}_{II}(\zeta_{\pi(M)}-\eta_1) \\
  \vdots &  & \\
  \mathbb{S}_{II}(\zeta_{\pi(1+\frac{M-K}{2})}-\eta_{K}) & \ldots & \mathbb{S}_{II}(\zeta_{\pi(M)}-\eta_K) \\
\mathbb{S}_{III}(\zeta_{\pi(1)}-\zeta_{\pi(1+\frac{M-K}{2})}), & \ldots & \mathbb{S}_{III}(\zeta_{\pi(1)}-\zeta_{\pi(M)} \\
  \vdots &  & \\
 \mathbb{ S}_{III}(\zeta_{\pi(\frac{M-K}{2})}-\zeta_{\pi(1+\frac{M-K}{2})}), & \ldots & \mathbb{S}_{III}(\zeta_{\pi(\frac{M-K}{2})}-\zeta_{\pi(M)}) \\
\end{array}
\right| \nonumber
\end{equation}
\begin{flushleft}\end{flushleft}
\emph{2.6 The average of the resolvent}
\begin{flushleft}\end{flushleft}\nopagebreak[3]
An interesting observation is that
 the non-universal functions emerging
in above expressions for the correlations of characteristic
polynomials can be expressed in terms of the large $N$ limit of
the averaged resolvent defined as
\begin{equation}
R^+_N(x)=\left\langle\mbox{Tr}\frac{1}{x-H}\right\rangle_H
\end{equation}
Here we assume that the parameter $x$ has an infinitesimal
positive imaginary part. It is then straightforward to observe that
$R^+_N(x)$ is expressed in terms of the
correlation function $\F_{II}^{K=1}(\epsilon,\mu)$. We have therefore
the following expression for the large $N$ limit of the averaged
resolvent:
\begin{equation}
R^+_N(x)=i\pi N\rho(x)-\frac{NV'(x)}{2}
\end{equation}
attributing a particular meaning to the non-universal factors:
\begin{equation}
N\rho(x)=\frac{1}{\pi}\Im m\left[R^+_N(x)\right],\; \alpha(x)=
-\pi\;\frac{\Re e\left[R^+_N(x)\right]}{\Im
m\left[R^+_N(x)\right]}
\end{equation}
\begin{flushleft}\end{flushleft}
\emph{2.7 Universality of $\F_{II}^K(\epsilon,\mu)$ at the center
of the spectrum}
\begin{flushleft}\end{flushleft}\nopagebreak[3]
 Equation (\ref{DysonLimitFII}) implies that the
following theorem is valid
\begin{thm}Assume that $\alpha(x)\equiv V'(x)/2\rho(x)=0$ (the center of the spectrum,
for example), where $V(x)$ is the potential function, $\rho(x)$ is
the density of states and $x$ belongs to the bulk of the spectrum.
Then correlation functions of \emph{ratios} of characteristic
polynomials of random Hermitian matrices are universal in the
Dyson scaling limit.
\end{thm}
\begin{flushleft}\end{flushleft}
\emph{Remark}. It can be observed that for the Gaussian case
$V(x)=x^2/2$ our results for the correlation functions (at the
centre of the spectrum $x=0$) are reduced to the formula
 by Andreev and Simons \cite{AS}.
A detailed derivation of the large $N$ asymptotics for the Gaussian
case can be found in \cite{II}.
\begin{flushleft}\end{flushleft}
\emph{2.8 Negative moments of characteristic polynomials}
\begin{flushleft}\end{flushleft}\nopagebreak[3]
 \emph{Positive} moments of characteristic
polynomials are determined by the following asymptotic expression
(Brezin and Hikami \cite{brezin1}):
\\
\begin{equation}
\left \langle \Z_N^{2K}[x,H] \right\rangle_H=
 [c_N]^{2K}
e^{KNV(x)}\left[N\rho(x)\right]^{K^2}\Upsilon_K^+
\end{equation}
\\
where
\\
\begin{equation}
\Upsilon_K^+=\prod_{l=0}^{K-1}l!/(l+K)!
\end{equation}
\\
 $\Upsilon_K^+$ is a
universal coefficient which also appears in positive moments of
zeta-function.

In the present paper we, in particular, obtain an asymptotic result for the
\emph{negative} moments of characteristic polynomials
$\M_{x,N}^K(\delta)$.
 These moments are defined as
\begin{equation}\label{NegativeMomentsDefinition}
\M_{x,N}^K(\delta)= \left \langle \Z_N^{-K}(x^+,H)\Z_N^{-K}(x^-,H)
\right\rangle_H
\end{equation}
where $x^{\pm}=x\pm\frac{i\delta}{2N\rho(x)}$, $x$ and $\delta$
are real parameters and $\rho(x)$ is the density of states.
For the large $N$ limit  we find that
the negative moments behave asymptotically as
\begin{equation}\label{NegativeMomentAsymptotic}
\M_{x,N}^K(\delta)=\left[2\pi\right]^K\left[c_N\right]^{-2K}
e^{-KNV(x)}\left[\frac{N\rho(x)}{\delta}\right]^{K^2}
\end{equation}
 Formula
(\ref{NegativeMomentAsymptotic}) should be compared with Gonek's
conjecture \cite{gonek} for the negative moments of Riemann zeta
function, which states that
\begin{equation}\label{Negativemomentsofzetafunction}
\lim\limits_{T\rightarrow\infty} \left\{\frac{1}{T}\int\limits_1^T
\left|\;\zeta\left(\frac{1}{2}+\frac{\delta}{\log
T}+it\right)\right|^{-2K}dt\right\}\sim\left(\frac{\log
T}{\delta}\right)^{K^2}
\end{equation}
A similarity between (\ref{NegativeMomentAsymptotic}) and
(\ref{Negativemomentsofzetafunction}) becomes apparent if we put
\begin{equation}
N\rho(x)=\log T
\end{equation}
in accord with the known expression for the mean density of Riemann zeroes.
 The
pre-factor $\left[2\pi\right]^K\left[c_N\right]^{-2K}$ in front of the
exponent in equation (\ref{NegativeMomentAsymptotic}) is
\emph{not} \emph{universal}, and as such it is irrelevant for the
comparison with the moments of zeta function  (e.g. for the
Gaussian case it is equal to $e^{NK}$ in the large $N$ limit).
On the other hand, the analogue of the universal coefficient $\Upsilon_K^+$
 for the negative moments is given by
\begin{equation}
\Upsilon_K^-=1
\end{equation}
as is immediately evident from the
 formula (\ref{NegativeMomentAsymptotic}). In
other words, \emph{the universal coefficient for negative moments
of characteristic polynomials which should also appear in the
negative moments of zeta function is 1.}

 Furthermore, we can see from the expression (\ref{NegativeMomentAsymptotic})
that the negative
moments diverge at $\delta\rightarrow 0$ as $\delta^{-\nu(K)}$,
with the exponent $\nu(K)$ being equal to $K^2$.
This fact fully agrees with the behaviour
conjectured by Berry
and Keating \cite{berry} for all unitary invariant
ensembles of $\beta=2$ symmetry class (see the
discussion in the Introduction).
\begin{flushleft}\end{flushleft}
\section{\textbf{Lagrange Interpolation Formula and Identities for
Characteristic Polynomials}}
\begin{flushleft}\end{flushleft}\nopagebreak[3]
 In this section we discuss some
consequences of the Lagrange interpolation formula (see
Szeg$\ddot{\mbox{o}}$ \cite{szego}). The obtained
relations enable us to derive
exact expressions for the correlation functions
(\ref{CorrelationOfProducts})-(\ref{CorrelationOfNegative}),
(\ref{CorrelationOfRatiosL}), (\ref{CorrelationOfRatiosS}).

Let $x_1,x_2,\ldots ,x_N$ be eigenvalues of the matrix $H$. Let us
associate with the
characteristic polynomial $\Z_{N}[\epsilon, H]$ of the matrix $H$
\begin{equation}\label{CharacteristicPolynomial}
\Z_N[\epsilon,H]=(\epsilon-x_1)(\epsilon-x_2)\ldots (\epsilon-x_N)
\end{equation}
 the fundamental polynomials of the Lagrange
interpolation:
\begin{equation}\label{l}
l_{\nu}(\epsilon)=\frac{\Z_N[\epsilon,H]}{\Z_N'[x_{\nu},H](\epsilon-x_{\nu})},
\;\;\nu=1,2,\ldots ,N
\end{equation}
 From equation
(\ref{CharacteristicPolynomial}) It is easy to observe that
\begin{equation}\label{CharacteristicPolynomiald}
\Z_N'[x_{\nu},H]=\prod\limits_{j\neq\nu}(x_{\nu}-x_j),
\;\;\nu=1,2,\ldots, N
\end{equation}
In particular, the equations (\ref{l}) and
(\ref{CharacteristicPolynomiald}) imply that
 the fundamental polynomials of the Lagrange
interpolation have the followin property:
\begin{equation}
l_{\nu}(x_{\mu})=\delta_{\nu\mu}\,\,.
\end{equation}
As each polynomial $P(x)$ of degree $N-1$ is determined uniquely
by its value in $N$ points, we have
\begin{equation}\label{PFROMX}
P(x)=P(x_1)l_1(x)+P(x_2)l_2(x)+\ldots +P(x_N)l_N(x)
\end{equation}
From the expression (\ref{PFROMX}) it follows that
\begin{eqnarray}
1=\sum\limits_{\nu=1}^{N}l_{\nu}(\epsilon)\qquad\nonumber\\
\epsilon=\sum\limits_{\nu=1}^{N}x_{\nu}l_{\nu}(\epsilon)\;\;\;\quad
\\
\ldots\qquad\qquad\quad\nonumber\\
\epsilon^{N-1}=\sum\limits_{\nu=1}^{N}x^{N-1}_{\nu}l_{\nu}(\epsilon)\nonumber
\end{eqnarray}
We immediately
conclude from the above expressions and the equation (\ref{l})
that the following algebraic identities must hold:
\begin{equation}
\sum\limits_{\nu=1}^{N}x_{\nu}^K/\Z_N'[x_{\nu},H]=0,\;\;\;0\leq
K\leq N-2
\end{equation}
and
\begin{equation}\label{FormulaForSimplificatios}
\frac{\epsilon^K}{\Z_N[\epsilon,H]}=
\sum\limits_{\nu=1}^{N}\frac{x_{\nu}^K}{\epsilon-x_{\nu}}\;
\frac{1}{\Z_N'[x_{\nu},H]},\;\;\; \forall\; K=0,\ldots , N-1
\end{equation}
With these equations in mind  it is not difficult to obtain a
representation for the Cauchy transform $h_{N-1}(\epsilon)$ of
monic orthogonal polynomial $\pi_{N-1}(x)$ defined by formula
(\ref{CauchyTransform}) in terms of a multi-variable integral
(for a derivation see \cite{strahov1}):
\begin{equation}
\gamma_{N-1}h_{N-1}(\epsilon)\nonumber
\end{equation}

\begin{equation}\label{IntegralRepresentofCauchyTransform}
=\frac{1}{Z_N}\int\prod\limits_{j=1}^{N}(\epsilon
-x_j)^{-1}\;e^{-N\sum\limits_{j=1}^{N}V(x_j)}\triangle^2(x_1,\ldots
, x_N)dx_1\ldots x_N
\end{equation}
The right hand side of this formula  can be looked at as the
average of $\Z^{-1}_N(\epsilon, H)$ taken over the ensemble of
unitary invariant Hermitian matrices. In other words, equation
(\ref{IntegralRepresentofCauchyTransform}) implies that
\begin{equation}
\left\langle\Z_N^{-1}[\epsilon,H]\right\rangle_H=\gamma_{N-1}h_{N-1}(\epsilon)
\end{equation}
In what follows we  compute more complicated correlation functions
of characteristic polynomials. Those correlation functions include
products of characteristic polynomials both in the numerator and the
denominator.
 The algebraic identity which  enables us to average the product of
the characteristic polynomials in the denominator is
\begin{equation}
\prod\limits_{l=1}^{M}\frac{\epsilon_l^{N-M}}{{\mathcal{Z}}_N[\epsilon_l,H]}\nonumber
\end{equation}

\begin{equation}\label{AlgebraicIdentity}
=\sum\limits_{\sigma} \left(
\prod\limits_{i,j=1}^M\frac{x_{\sigma(i)}^{N-M}}{\epsilon_{\sigma(j)}-
x_{\sigma(i)}}\right)\; \frac{\triangle(x_{\sigma(1)},\ldots
,x_{\sigma(M)})\triangle(x_{\sigma(M+1)},\ldots
,x_{\sigma(N)})}{\triangle(x_{\sigma(1)},\ldots ,x_{\sigma(N)})}
\end{equation}
where $\sigma\in {\sf{S}}_N/{\sf{S}}_{N-M}\times {\sf{S}}_M$,
${\sf{S}}_N$ is the permutation group of the full index set $1,\ldots
,N$, whereas ${\sf{S}}_M$ is the permutation group of the first $M$
indices and ${\sf{S}}_{N-M}$ is the permutation group of the
remaining $N-M$ indices.  Identity (\ref{AlgebraicIdentity}) was
proved in \cite{strahov1} and follows as a consequence of the
Cauchy-Littlewood formula \cite{schur}
\begin{equation}\label{Cauchy-Littlewood formula}
\prod\limits_{j=1}^{M}\prod\limits_{i=1}^{N}(1-x_iy_j)^{-1}=
\sum\limits_{\lambda}s_{\lambda}(x_1,\ldots
,x_N)s_{\lambda}(y_1,\ldots ,y_M)
\end{equation}
and the Jacobi-Trudi identity \cite{schur}:
\begin{equation}
s_{\lambda}(x_1,\ldots,x_N)=
\frac{\mbox{det}\left(x_i^{\lambda_j-j+N}\right)
}{\triangle(x_1,\ldots ,x_N)}
\end{equation}
where the Schur polynomial $s_{\lambda}(x_1,\ldots ,x_N)$
corresponds to a partition $\lambda$, and the indices $i,j$ take
their values from $1$ to $N$.
\begin{flushleft}\end{flushleft}
\section{\textbf{Finite Correlation Functions}}
\begin{flushleft}\end{flushleft}\nopagebreak[3]
\emph{4.1 Correlation function
$\F_{I}^K(\hat\lambda,\hat\mu)=\left\langle\prod\limits_{1}^{K}\Z_N\left[\lambda_j,H\right]
\Z_N\left[\mu_j, H\right] \right\rangle_{H}$}
\begin{flushleft}\end{flushleft}\nopagebreak[3]
The correlation function of products of characteristic polynomials
$\F_{I}^K(\hat\lambda,\hat\mu)$ was investigated in detail by
Brezin and Hikami \cite{brezin1,brezin2}. For finite size $N$
these authors have demonstrated that the correlation function could be
rewritten in a determinant form. Namely,
\begin{equation}\label{BrezinDeterminantPolynomials}
\F_{I}^K(\hat\lambda,\hat\mu)
=\frac{1}{\triangle(\hat\lambda,\hat\mu)}\; \mbox{det}\left|
\begin{array}{cccc}
  \pi_N(\lambda_1) & \pi_{N+1}(\lambda_1)
   & \ldots & \pi_{N+2K-1}(\lambda_1) \\
  \pi_N(\lambda_2) & \pi_{N+1}(\lambda_2)
   & \ldots & \pi_{N+2K-1}(\lambda_2) \\
  \vdots &  &  &  \\
  \pi_N(\lambda_K) & \pi_{N+1}(\lambda_K)
   & \ldots & \pi_{N+2K-1}(\lambda_K) \\
  \pi_N(\mu_1) & \pi_{N+1}(\mu_1) & \ldots
   & \pi_{N+2K-1}(\mu_1) \\
  \pi_N(\mu_2) & \pi_{N+1}(\mu_2)
   & \ldots & \pi_{N+2K-1}(\mu_2) \\
  \vdots &  &  &  \\
  \pi_N(\mu_K) & \pi_{N+1}(\mu_K)
   & \ldots & \pi_{N+2K-1}(\mu_K)
\end{array}\right|
\end{equation}
The same correlation function $\F_{I}^K(\hat\lambda,\hat\mu)$ has
also an alternative representation
in terms of a determinant of a kernel constructed from monic
orthogonal polynomials.
\begin{flushleft}\end{flushleft}
\begin{prop}
(\cite{brezin1,brezin2}): The correlation function of products of
characteristic polynomials is governed by a two-point  kernel
function constructed from monic orthogonal polynomials,
\begin{equation}\label{BrezinDeterminantKernel}
\F_{I}^K(\hat\lambda, \hat\mu)=
\frac{C_{N,K}}{\triangle(\hat\lambda)\triangle(\hat\mu)}\;
\mbox{det} \left[ W_{I,N+K}(\lambda_i,\mu_j)\right]_{1\leq i,j\leq
K}
\end{equation}
where the kernel $W_{I,N+K}(x,y)$ is given by the formula
\begin{equation}\label{KernelWI}
W_{I,N+K}^{K}(\lambda,\mu)=
\frac{\pi_{N+K}(\lambda)\pi_{N+K-1}(\mu)-
\pi_{N+K}(\mu)\pi_{N+K-1}(\lambda)}{\lambda-\mu}
\end{equation}
The constant $C_{N,K}$ can be expressed in terms of the
coefficients $\gamma_l$ defined by equation (\ref{gamma})
\begin{equation}\label{ConstantCN,K}
C_{N,K}=[c_{N+K-1}]^{-2K}\;\prod\limits_{N}^{N+K-1}(c_l)^2= \frac{
\left[\gamma_{N+K-1}\right]^K}{\prod_N^{N+K-1}\gamma_l}
\end{equation}
\end{prop}
\begin{flushleft}\end{flushleft}
\begin{proof}
To prove formula (\ref{BrezinDeterminantKernel}) we observe that
the correlation function $\F_{I}^K(\hat\lambda, \hat\mu)$ can be
represented as the integral
\begin{equation}
\F_{I}^K(\hat\lambda,
\hat\mu)=\frac{Z^{-1}_N}{\triangle(\hat\lambda)\triangle(\hat\mu)}\;
\int d^N\hat x\;
e^{-N\sum\limits_{i=1}^{N}V(x_i)}\triangle(\hat\lambda,\hat
x)\triangle(\hat\mu,\hat x)
\end{equation}
This integral can be evaluated using the method of orthogonal
polynomials. Namely, we rewrite the Vandermonde determinants as
determinants of monic orthogonal polynomials. Then the product of
the Vandermonde determinants in the integrand above can be rewritten
as a sum over permutations, i.e.
\begin{equation}
\triangle(\hat\lambda,\hat x)\triangle(\hat\mu,\hat x)\nonumber
\end{equation}

\begin{equation}
\qquad\qquad= \sum\limits_{\sigma, \rho\;\in\;
{\textsf{S}}_{N+K}}(-)^{\nu_{\sigma}+\nu_{\rho}}
[\pi_{\sigma(1)-1}(\lambda_1)\ldots\pi_{\sigma(K)-1}(\lambda_K)\nonumber
\end{equation}

\begin{equation}
\qquad\qquad\;\;\;\qquad\qquad\quad\times\;
\qquad\;\;\pi_{\sigma(K+1)-1}(x_1)\ldots\pi_{\sigma(N+K)-1}(x_N)
\end{equation}

\begin{equation}
\qquad\qquad\;\;\;\qquad\qquad\quad\times
\;\qquad\;\;\pi_{\rho(1)-1}(\mu_1)\ldots\pi_{\rho(K)-1}(\mu_K)\nonumber
\end{equation}

\begin{equation}
\qquad\qquad\;\;\;\qquad\qquad\quad\times
\;\qquad\;\;\pi_{\rho(K+1)-1}(x_1)\ldots\pi_{\rho(N+K)-1}(x_N)]\nonumber
\end{equation}
We insert the above formula into the integrand of
$\F_{I}^K(\hat\lambda, \hat\mu)$ and integrate over the variables
$x_1,\ldots ,x_N$. The orthogonality of monic polynomials leads to
the expression
\begin{equation}
\F_{I}^K(\hat\lambda,
\hat\mu)=\frac{Z^{-1}_N\left[\prod\limits_0^{N+K-1}
c_j^2\right]}{\triangle(\hat\lambda)\triangle(\hat\mu)}\;\nonumber
\end{equation}

\begin{equation}
\qquad\qquad\;\;\times\; \sum\limits_{\sigma, \rho\;\in\;
{\textsf{S}}_{N+K}}(-)^{\nu_{\sigma}+\nu_{\rho}}
[q_{\sigma(1)-1}(\lambda_1)\ldots q_{\sigma(K)-1}(\lambda_K)
\nonumber
\end{equation}

\begin{equation}
\qquad\qquad\qquad\qquad\;\;\qquad\times\qquad
\;\;\;\;q_{\rho(1)-1}(\mu_1)\ldots q_{\rho(K)-1}(\mu_K)\;
\end{equation}

\begin{equation}
\qquad\qquad\qquad\qquad\;\;\qquad\times\qquad
\;\;\;\;\delta_{\sigma(K+1)\rho(K+1)}\ldots
\delta_{\sigma(K+N)\rho(K+N)}] \nonumber
\end{equation}
Here we introduced the  polynomials $q_l(x)=c_l^{-1}\pi_l(x)$
normalized with
respect to the measure $d\mu(x)=\exp\left(-NV(x)\right)$.
The sum in the equation above can be further
transformed to a determinant (see
Appendix A) and we end up with the following expression
\begin{equation}\label{001}
\F_{I}^K(\hat\lambda, \hat\mu)=
\frac{N!\left[\prod\limits_0^{N+K-1}
c_j^2\right]}{Z_N\triangle(\hat\lambda)\triangle(\hat\mu)}\;
\mbox{det}\left[\sum\limits_{0}^{N+K-1}q_l(\lambda_i)q_l(\mu_j)
\right]_{1\leq i,j\leq K}
\end{equation}
Applying to (\ref{001}) the Christoffel-Darboux formula (see, for
example, Szeg\"{o} \cite{szego}) we recover the expression
(\ref{BrezinDeterminantKernel}).
\end{proof}
\begin{flushleft}\end{flushleft}
\emph{4.2 Correlation function $\F_{II}^K(\hat\epsilon,
\hat\mu)=\left\langle \prod\limits_{1}^{K}
\frac{\Z_N\left[\mu_j,H\right]}{\Z_N\left[\epsilon_j,H\right]}
\right\rangle_{H}$}
\begin{flushleft}\end{flushleft}\nopagebreak[3]
 Here we derive an exact formula
representing $\F_{II}^K(\hat\mu, \hat\epsilon)$ as a determinant
of the kernel $W_{II,N}(\epsilon,\mu)$.
\begin{flushleft}\end{flushleft}
\begin{prop}\label{FII}
Let $\Im m \;\epsilon_j\neq 0$, $j=1,\ldots ,K$. Then the
correlation function of ratios of characteristic polynomials is
determined by a two-point kernel constructed from
monic orthogonal polynomials and their Cauchy transforms. More
precisely, the following formula holds
\begin{equation}
\F_{II}^K(\hat\epsilon,
\hat\mu)=(-)^{\frac{K(K-1)}{2}}\left[\gamma_{N-1}\right]^K
\frac{\triangle(\hat\epsilon,\hat\mu)}{\triangle^2(\hat\epsilon)\triangle^2(\hat\mu)}
\;\mbox{det}\left[W_{II,N}(\epsilon_i,\mu_j)\right]_{1\leq i,j\leq
K}
\end{equation}
where the kernel $W_{II,N}(\epsilon,\mu)$ is given by
\begin{equation}\label{WII}
W_{II,N}(\epsilon,\mu)=\frac{h_{N}(\epsilon)\pi_{N-1}(\mu)-h_{N-1}(\epsilon)\pi_{N}(\mu)
  }{\epsilon-\mu}
  \end{equation}
  and the constant $\gamma_{N-1}$ is defined by the equation
  (\ref{gamma}).
\end{prop}
\begin{flushleft}\end{flushleft}
\begin{proof}
We propose  a "reduction procedure". The idea is to reduce computation of
the correlation functions containing {\it ratios} of characteristic polynomials to
 that of the correlation function which contains only
{\it products} of characteristic polynomials. Namely, we exploit
the identity (\ref{AlgebraicIdentity}) to express the denominator,
$\prod_1^K\Z_N^{-1}[\epsilon_j,H]$, as a sum over permutations.
$\F_{II}^K(\hat\epsilon, \hat\mu)$ is a multi-variable integral
with the measure defined by the eigenvalue density function
(\ref{EigenvalueDensity}), so the integrand is symmetric under
permutations of the variables of the integration. (Recall that
$x_1,\ldots ,x_N$ denote eigenvalues of the Hermitian matrix $H$,
$\mbox{dim}\; H=N$). It means that each permutation gives the
same contribution to the correlation function. The total number of
those permutations is $\frac{N!}{(N-K)!K!}$. We then find that
\begin{equation}
\F_{II}^K(\hat\epsilon, \hat\mu)= \frac{N!}{(N-K)!K!}
\left[\prod\limits_{1}^K\epsilon_l^{K-N}\right]
\end{equation}

\begin{equation}
\qquad\qquad\times\; \left\langle\left[\prod\limits_{i,j=1}^K
 \frac{x_i^{N-K}(\mu_j-x_i)}{\epsilon_j-x_i}\right]
 \left[\prod\limits_{l=1}^K\prod_{s=K+1}^{N}\frac{\mu_l-x_s}{x_l-x_s}
 \right]\right\rangle_H\nonumber
 \end{equation}
 The next step is to decompose the integration measure
 in accordance with the following expression for the eigenvalue
 density function
 \begin{equation}
 {\it{P}}^{(N)}(x_1,\ldots
,x_N)=\frac{Z_{N-K}Z_K}{Z_N}
\left[\prod\limits_{l=1}^{K}\prod\limits_{s=K+1}^{N} (x_l-x_s)^2
\right]\nonumber
\end{equation}

\begin{equation}
 \qquad\qquad\qquad\qquad\times\; {\it{P}}^{(K)}(x_1,\ldots
,x_K){\it{P}}^{(N-K)}(x_{K+1},\ldots ,x_{N})
\end{equation}
which allows one to rewrite the correlation function as

\begin{equation}
\F_{II}^K(\hat\epsilon, \hat\mu)=\frac{N!}{(N-K)!K!}
\frac{Z_{N-K}}{Z_N}\left[\prod\limits_1^K\epsilon_l^{K-N}
\right]\nonumber
\end{equation}

 \begin{equation}\label{KdimIntegral}
 \;\qquad\qquad\times\;\;\int dx_1\ldots
dx_K\;\triangle^2(x_1,\ldots ,x_K)
\end{equation}

\begin{equation}
 \;\qquad\qquad\times\left[
\prod\limits_{i,j=1}^K\frac{e^{-NV(x_i)}x_i^{N-K}(\mu_j-x_i)}{\epsilon_j-x_i}
\right] \left\langle\prod\limits_1^K\Z_{N-K}[\mu_l,\tilde
H]\Z_{N-K}[x_l,\tilde H]\right\rangle_{\tilde H}\nonumber
\end{equation}
Here $\mbox{dim}\; \tilde H=N-K, \tilde H^{\dag}=\tilde H$. We
then observe that the original integration over $N$ variables is
replaced by an integration over $K$ variables. Moreover, we
notice that the
correlation function of products of characteristic polynomials
emerges in the integrand of the formula (\ref{KdimIntegral}). Then
the equation
(\ref{BrezinDeterminantKernel}) yields
\begin{equation}
\left\langle\prod\limits_1^K\Z_{N-K}[\mu_l,\tilde
H]\Z_{N-K}[x_l,\tilde H]\right\rangle_{\tilde H}=
\end{equation}

\begin{equation}
\qquad\qquad\frac{C_{N-K,K}}{\triangle(x_1,\ldots
,x_K)\triangle(\mu_1,\ldots
,\mu_K)}\;\mbox{det}\left[W_{I,N}(\mu_i,x_j)\right]_{1\leq i,j
\leq K}\nonumber
\end{equation}
which leads to essential simplifications in the expression for
$\F_{II}^K(\hat\epsilon, \hat\mu)$:
\begin{equation}
\F_{II}^K(\hat\epsilon, \hat\mu)=C_{N-K,K}\;\frac{N!}{(N-K)!K!}
\frac{Z_{N-K}}{Z_N}\left[\prod\limits_1^K\epsilon_l^{K-N}
/\triangle(\hat\mu)\right]\nonumber
\end{equation}

\begin{equation}
\qquad\qquad\times\;\; \int\ dx_1\ldots dx_K\;\triangle(x_1,\ldots
,x_K)
\end{equation}

\begin{equation}
\qquad\qquad\times\left[
\prod\limits_{i,j=1}^K\frac{e^{-NV(x)}x_i^{N-K}(\mu_j-x_i)}{\epsilon_j-x_i}
\right]\mbox{det}\left[W_{I,N}(\mu_i,x_j)\right]_{1\leq i,j\leq K}
\nonumber
\end{equation}
The first two terms in the integrand can be further rewritten as
$K\times K$ determinant,
\begin{equation}
\triangle(x_1,\ldots ,x_K) \left[
\prod\limits_{i,j=1}^K\frac{e^{-NV(x)}x_i^{N-K}(\mu_j-x_i)}{\epsilon_j-x_i}
\right]\nonumber
\end{equation}

\begin{equation}
\qquad\qquad\qquad\qquad=(-)^{\frac{K(K-1)}{2}}\mbox{det}\left[f_i(x_j)\right]_{1\leq
i,j\leq K}
\end{equation}
where
\begin{equation}
f_i(x)=x^{N-K+i-1}e^{-NV(x)}\left[\prod\limits_1^K\frac{\mu_l-x}{\epsilon_l-x}
\right]\equiv \frac{x^{i-1}g(x)}{\prod_1^K(\epsilon_l-x)}
\end{equation}
i.e.
\begin{equation}
g(x)=x^{N-K}e^{-NV(x)}\prod\limits_1^K(\mu_l-x)
\end{equation}
Now we simplify $\mbox{det}\left[f_i(x_j)\right]$. In order to do
this we notice that from
(\ref{FormulaForSimplificatios}) follows that,
\begin{equation}
x^{i-1}\prod\limits_1^K\frac{g(x)}{\epsilon_l-x} =
(-)^K\sum\limits_{\nu=1}^K\frac{\epsilon_{\nu}^{i-1}}{x-\epsilon_{\nu}}
\;\frac{g(x)}{\prod_{l\neq\nu}(\epsilon_{\nu}-\epsilon_{l})}\nonumber
\end{equation}
and thus we can write
\begin{equation}
\mbox{det}\left[f_i(x_j)\right]_{1\leq i,j\leq K} = (-)^{K^2}
\mbox{det}\left[\sum\limits_{\nu=1}^K\frac{\epsilon_{\nu}^{i-1}}{x_j-\epsilon_{\nu}}
\;\frac{g(x_j)}{\prod_{l\neq\nu}(\epsilon_{\nu}-\epsilon_{l})}\right]_{1\leq
i,j\leq K}\nonumber
\end{equation}

\begin{equation}
=(-)^{K^2}\mbox{det}\left(\epsilon_{\nu}^{i-1}\right)\mbox{det}
\left[ \frac{g(x_j)}{x_j-\epsilon_{\nu}}
\;\frac{1}{\prod_{l\neq\nu}(\epsilon_{\nu}-\epsilon_{l})}\right]_{1\leq
j,\nu\leq K}
\end{equation}

\begin{equation}
=(-)^{K^2}(-)^{K(K-1)} \frac{1}{\triangle(\hat\epsilon)}\;
\mbox{det} \left[ \frac{g(x_j)}{x_j-\epsilon_{\nu}}
\;\right]_{1\leq j,\nu\leq K}\nonumber
\end{equation}
Therefore,
\begin{equation}
\F_{II}^K(\hat\epsilon,
\hat\mu)=(-)^{K+1}C_{N-K,K}\;\frac{N!}{(N-K)!K!}
\frac{Z_{N-K}}{Z_N}\;\frac{\prod_1^K\epsilon_l^{K-N}}{
\triangle(\hat\mu)\triangle(\hat\epsilon)}\nonumber
\end{equation}

\begin{equation}\label{FIILast}
\times\;\;\int dx_1\ldots dx_K \;\;\mbox{det}
\left[\frac{g(x_j)}{x_j-\epsilon_{\nu}}\;\right]_{1\leq i,j\leq K}
\mbox{det}\left[W_{I,N}(\mu_i,x_j)\right]_{1\leq i,j\leq K}
\end{equation}
Now it is not difficult to calculate the last integral. Let us
rewrite determinants as sums over permutations
\begin{equation}
\int dx_1\ldots dx_K \;\;\mbox{det}
\left[\frac{g(x_j)}{x_j-\epsilon_{\nu}}\;\right]_{1\leq i,j\leq K}
\mbox{det}\left[W_{I,N}(\mu_i,x_j)\right]_{1\leq i,j\leq
K}\nonumber
\end{equation}

\begin{equation}
=\sum\limits_{\sigma,\;\rho\; \in\;
{\textsf{S}}_K}(-)^{\nu_{\sigma}+\nu_{\rho}}\int dx_1
\left[\frac{g(x_1)}{x_1-\epsilon_{\sigma
(1)}}\right]W_{I,N}(\mu_{\rho(1)},x_1)
\end{equation}

\begin{equation}
\qquad\qquad\times\ldots\times \int dx_K
\left[\frac{g(x_K)}{x_K-\epsilon_{\sigma
(K)}}\right]W_{I,N}(\mu_{\rho(K)},x_K)\nonumber
\end{equation}
We compute the integrals above using orthogonality of monic
polynomials (for details  see equation
(\ref{CalculationoftheintegralW}), where a similar integral is
calculated). This yields
\begin{equation}
\int dx_1\ldots dx_K \;\;\mbox{det}
\left[\frac{g(x_j)}{x_j-\epsilon_{\nu}}\;\right]_{1\leq i,j\leq K}
\mbox{det}\left[W_{I,N}(\mu_i,x_j)\right]_{1\leq i,j\leq
K}\nonumber
\end{equation}

\begin{equation}
=\sum\limits_{\sigma,\;\rho\; \in\;
{\textsf{S}}_K}(-)^{\nu_{\sigma}+\nu_{\rho}}\epsilon^{N-K}_{\sigma(1)}
\left[\prod\limits_1^K\left(\mu_l-\epsilon_{\sigma(1)}\right)\right]
\;2\pi i\; W_{II,N}(\epsilon_{\sigma(1)},\mu_{\rho(1)})\nonumber
\end{equation}

\begin{equation}
\qquad\qquad\times\ldots \times\;\; \epsilon^{N-K}_{\sigma(K)}
\left[\prod\limits_1^K\left(\mu_l-\epsilon_{\sigma(K)}\right)\right]
\;2\pi i\; W_{II,N}(\epsilon_{\sigma(K)},\mu_{\rho(K)})
\end{equation}

\begin{equation}
=(2\pi i)^K K!\left[\prod\limits_1^K\epsilon_l^{N-K}\right]
\left[\prod\limits_{i,j=1}^K(\mu_i-\epsilon_j)\right] \mbox{det}
\left[W_{II,N}(\epsilon_i,\mu_j)\right]_{1\leq i,j\leq K}\nonumber
\end{equation}
We insert the obtained expression  to (\ref{FIILast}) and with
some simple algebra prove the proposition.
\end{proof}
\begin{flushleft}\end{flushleft}
\emph{4.3 Correlation function $\F_{III}^K(\hat\varpi,
\hat\omega)=\left\langle \prod\limits_{1}^{K}
\frac{1}{\Z_N\left[\varpi_j,H\right]\Z_N\left[\omega_j,H\right]}
\right\rangle_{H}$}
\begin{flushleft}\end{flushleft}\nopagebreak[3]
Similar to the previous cases, the correlation functions that
contain an even number of characteristic polynomials in the
denominator are governed by a (different) two-point kernel.
However, such correlation
functions are not readily expressible as a determinant of a kernel divided
by two Vandermonde determinants. Now an exact formula will be
slightly more complicated. In fact, we prove that the
 correlation function $\F_{III}^K(\hat\varpi, \hat\omega)$ is a
sum over permutations.
\begin{flushleft}\end{flushleft}
\begin{prop}
Define $2K$ dimensional vector $\hat\epsilon$,
\begin{equation}
\hat\epsilon=\left(\hat\varpi,\hat\omega\right),\;\;\;\Im
m\;\epsilon_j\neq 0
\end{equation}
Then the correlation function which contains an even number of
characteristic polynomials in the denominator,
$\F_{III}^K(\hat\epsilon)\equiv\F_{III}^K(\hat\varpi,
\hat\omega)$, can be expressed as the following sum over
permutations
\begin{equation}
\F_{III}^K(\hat\epsilon)
\end{equation}

\begin{equation}
=(-)^K\;\frac{\left[\gamma_{N-1}\right]^{2K}}{(2K)!}
\sum\limits_{\pi\;\in\;
{\textsf{S}}_{2K}}\frac{\mbox{det}\left[W_{III,N-K}(\epsilon_{\pi(i)},
\epsilon_{\pi(K+j)})\right]_{1\leq i,j\leq K}
}{\triangle(\epsilon_{\pi(1)},\ldots
,\epsilon_{\pi(K)})\triangle(\epsilon_{\pi(K+1)},\ldots
,\epsilon_{\pi(2K)})}\nonumber
\end{equation}
 The two-point kernel $W_{III,N-K}(\epsilon,\omega)$ is
 constructed from the Cauchy transforms of monic orthogonal
 polynomials,
 \begin{equation}\label{WIII}
W_{III,N-K}(\epsilon,\omega)=\frac{h_{N-K}(\epsilon)h_{N-K-1}(\omega)-
h_{N-K-1}(\epsilon)h_{N-K}(\omega)}{\epsilon-\omega}
\end{equation}
The constant $\gamma_{N-1}$ is determined by equation
(\ref{gamma}).
\end{prop}
\begin{flushleft}\end{flushleft}
\begin{proof}
We follow the procedure applied previously to the correlation
function $\F_{II}^K(\hat\epsilon,\hat\mu)$. Instead of equation
(\ref{FIILast}) we obtain
\begin{equation}
\F_{III}^K(\hat\epsilon)=(-)^K\;C_{N-2K,K}\;\frac{N!}{(N-2K)!(2K)!}
\frac{Z_{N-2K}}{Z_N}\;\frac{\prod_1^K\epsilon_l^{2K-N}}{
\triangle(\hat\epsilon)}\nonumber
\end{equation}

\begin{equation}\label{FIIILast}
\times\;\int dx_1\ldots dx_{2K}
\left[\prod\limits_{s=1}^K\prod\limits_{l=K+1}^{2K}(x_s-x_l)\right]
\end{equation}

\begin{equation}
\times\;\;\;\;\;\mbox{det}
\left[\frac{x_i^{N-2K}e^{-NV(x_i)}}{x_i-\epsilon_{j}}\;\right]_{1\leq
i,j\leq 2K} \;
\mbox{det}\left[W_{I,N-K}(x_i,x_{K+j})\right]_{1\leq i,j\leq K}
\nonumber
\end{equation}
To compute the integral above we proceed as follows
\begin{equation}
\int dx_1\ldots
dx_{2K}\left[\prod\limits_{s=1}^K\prod\limits_{l=K+1}^{2K}(x_s-x_l)\right]
\nonumber
\end{equation}

\begin{equation}
\times\;\;\;\; \; \mbox{det}
\left[\frac{x_i^{N-2K}e^{-NV(x_i)}}{x_i-\epsilon_{j}}\;\right]_{1\leq
i,j\leq 2K} \;
\mbox{det}\left[W_{I,N-K}(x_i,x_{K+j})\right]_{1\leq i,j\leq
K}\nonumber
\end{equation}

\begin{equation}
=\sum\limits_{\pi\;\in\; {\textsf{S}}_{2K}}(-)^{\nu_{\pi}}
\sum\limits_{\sigma\;\in\;{\textsf{S}}_{K}}(-)^{\nu_{\sigma}}\nonumber
\end{equation}

\begin{equation}
\times \int dx_1\ldots
dx_K\left[\frac{x_1^{N-2K}e^{-NV(x_1)}}{\epsilon_{\pi(1)}-x_1}\right]\times\ldots
\times\left[\frac{x_K^{N-2K}e^{-NV(x_K)}}{\epsilon_{\pi(K)}-x_K}\right]\nonumber
\end{equation}

\begin{equation}
\times\;\int
dy_1\frac{y_1^{N-2K}e^{-NV(y_1)}\prod_1^K(x_l-y_1)}{\epsilon_{\pi(K)}-y_1}
\;W_{I,N-K}(x_{\sigma(1)},y_1)\nonumber
\end{equation}

\begin{equation}
\times\ldots\times\nonumber
\end{equation}

\begin{equation}
\times\; \int
dy_K\;\left[\frac{y_K^{N-2K}e^{-NV(y_K)}\prod_1^K(x_l-y_K)}{\epsilon_{\pi(2K)}-y_K}
\right]\;W_{I,N-K}(x_{\sigma(K)},y_K)\nonumber
\end{equation}
We have
\begin{equation}
\int
dy\;\left[\frac{y^{N-2K}e^{-NV(y)}\prod_1^K(x_l-y)}{\epsilon_{i}-y}\right]
\;W_{I,N-K}(x_j,y)\nonumber
\end{equation}

\begin{equation}
=\int dy\;\frac{y^{N-2K}e^{-NV(y)} \left[\prod_1^{j-1}(x_l-
y)\right]\left[\prod_{j+1}^K(x_l-y)\right]}{\epsilon_{i}-y}
\nonumber
\end{equation}

\begin{equation}
\times\;\;\left[\pi_{N-K}(x_j)\pi_{N-K-1}(y)-\pi_{N-K}(y)\pi_{N-K-1}(x_j)\right]
\nonumber
\end{equation}

\begin{equation}\label{CalculationoftheintegralW}
=\left[\prod_i^{j-1}(x_l-\epsilon_i)\right]\left[\prod_{j+1}^K(x_l-\epsilon_i)\right]
\end{equation}

\begin{equation}
\times\;\int dy\;\left[\frac{y^{N-2K}e^{-NV(y)}
}{\epsilon_{i}-y}\right]
\;\left[\pi_{N-K}(x_j)\pi_{N-K-1}(y)-\pi_{N-K}(y)\pi_{N-K-1}(x_j)\right]
\nonumber
\end{equation}
where the orthogonality of the monic polynomials with respect to
the weight function $\exp\left[-NV(x)\right]$ is used. Therefore,
\begin{equation}
\int
dy\;\left[\frac{y^{N-2K}e^{-NV(y)}\prod_1^K(x_l-y)}{\epsilon_{i}-y}
\right]\;W_{I,N-K}(x_j,y)\nonumber
\end{equation}

\begin{equation}
 =\left(-2\pi
i\right)\epsilon_i^{N-2K}\left[\prod_1^K(x_l-\epsilon_i)\right]W_{II,
N-K}(\epsilon_i,x_j)\nonumber
\end{equation}
This yields
\begin{equation}
\int dx_1\ldots
dx_{2K}\left[\prod\limits_{s=1}^K\prod\limits_{l=K+1}^{2K}(x_s-x_l)\right]
\nonumber
\end{equation}

\begin{equation}
\times\;\;\;\; \; \mbox{det}
\left[\frac{x_i^{N-2K}e^{-NV(x_i)}}{x_i-\epsilon_{j}}\;\right]_{1\leq
i,j\leq 2K} \;
\mbox{det}\left[W_{I,N-K}(x_i,x_{K+j})\right]_{1\leq i,j\leq
K}\nonumber
\end{equation}

\begin{equation}
= \left(-2\pi i\right)^K\sum\limits_{\pi\;\in\;{\textsf{S}}_{2K}}
(-)^{\nu_{\pi}}\left[\prod\limits_{1}^K
\epsilon_{\pi(K+1)}^{N-2K}\right]\sum\limits_{\sigma\;\in\;{\textsf{S}}_K}
(-)^{\nu_{\sigma}}\nonumber
\end{equation}

\begin{equation}
\int
dx_{\sigma(1)}\frac{x_{\sigma(1)}^{N-2K}e^{-NV(x_{\sigma(1)})}
\left[\prod_1^K\left(x_{\sigma(1)}-\epsilon_{\pi(K+l)}\right)\right]
}{\epsilon_{\pi(\sigma(1))}-x_{\sigma(1)}}\;W_{II,
N-K}(\epsilon_{\pi(K+\sigma(1))},x_{\sigma(1)}) \nonumber
\end{equation}

\begin{equation}
\times\ldots\times \nonumber
\end{equation}

\begin{equation}
\int
dx_{\sigma(K)}\frac{x_{\sigma(K)}^{N-2K}e^{-NV(x_{\sigma(K)})}
\left[\prod_1^K\left(x_{\sigma(K)}-\epsilon_{\pi(K+l)}\right)\right]
}{\epsilon_{\pi(\sigma(K))}-x_{\sigma(K)}}\;W_{II,
N-K}(\epsilon_{\pi(K+\sigma(K))},x_{\sigma(K)}) \nonumber
\end{equation}

\begin{equation}
=(-2\pi i)^{2K}\sum\limits_{\pi\;\in\;{\textsf{S}}_{2K}}
(-)^{\nu_{\pi}}\left[\prod\limits_{1}^K
\epsilon_{\pi(K+1)}^{N-2K}\right]\sum\limits_{\sigma\;\in\;{\textsf{S}}_K}
(-)^{\nu_{\sigma}}\left[\prod\limits_{1}^K\epsilon_{\pi(\sigma(l))}^{N-2K}
\right]\nonumber
\end{equation}

\begin{equation}
\left[\prod_1^K\left(\epsilon_{\pi(\sigma(1))}-\epsilon_{\pi(K+l)}\right)\right]
W_{III,N-K}(\epsilon_{\pi(K+\sigma(1))},\epsilon_{\pi(\sigma(1))})
\nonumber
\end{equation}

\begin{equation}
\times\ldots\times \nonumber
\end{equation}

\begin{equation}
\left[\prod_1^K\left(\epsilon_{\pi(\sigma(K))}-\epsilon_{\pi(K+l)}\right)\right]
W_{III,N-K}(\epsilon_{\pi(K+\sigma(K))},\epsilon_{\pi(\sigma(K))})
\nonumber
\end{equation}

\begin{equation}
=(-2\pi i
)^{2K}\left[\prod\limits_{1}^{2K}\epsilon_l^{N-2K}\right]\nonumber
\end{equation}

\begin{equation}
\times\;\sum\limits_{\pi\;\in\;{\textsf{S}}_{2K}}(-)^{\nu_{\pi}}
\left[
\prod\limits_{i,j=1}^K\left(\epsilon_{\pi(i)}-\epsilon_{\pi(K+j)}\right)
\right]
\mbox{det}\left[W_{III,N-K}(\epsilon_{\pi(i)},\epsilon_{\pi(K+j)})
\right]_{1\leq i,j\leq K}\nonumber
\end{equation}

\begin{equation}
=(-2\pi i
)^{2K}\left[\prod\limits_{1}^{2K}\epsilon_l^{N-2K}\right]
\sum\limits_{\pi\;\in\;{\textsf{S}}_{2K}}
\frac{\mbox{det}\left[W_{III,N-K}(\epsilon_{\pi(i)},\epsilon_{\pi(K+j)})
\right]_{1\leq i,j\leq K}}{\triangle(\epsilon_{\pi(1)},\ldots
,\epsilon_{\pi(K)}) \triangle(\epsilon_{\pi(K+1)},\ldots
,\epsilon_{\pi(2K)})}\nonumber
\end{equation}
We insert the above expression to the formula (\ref{FIIILast}) and prove
the proposition.
\end{proof}
\begin{flushleft}\end{flushleft}
\emph{4.4 Correlation function $\F_{IV}^K(\hat\epsilon,
\hat\mu)=\left\langle \frac{\prod\limits_{1}^{K}
\Z_N\left[\mu_l,H\right]}{\prod\limits_1^M\Z_N\left[\epsilon_j,H\right]}
\right\rangle_{H},\;K>M$}
\begin{flushleft}\end{flushleft}\nopagebreak[3]
 As a result of the fact that the numbers
of characteristic polynomials in the denominator and the numerator are
not equal to each other, the correlation function of
characteristic polynomials turns out to be determined by two kernel
functions.
\begin{flushleft}\end{flushleft}
\begin{prop}
Let $\Im m\;\epsilon_j\neq 0$. For the correlation function which
contains $K$ characteristic polynomials in the numerator and $M<K$
characteristic polynomials in the denominator  the following
formula holds
\begin{equation}
\F_{IV}^K(\hat\epsilon, \hat\mu)=(-)^{\frac{M(K-1)}{2}}\;\;
\frac{\left[\gamma_{N-M+L-1}\right]^L}{\left[\prod_N^{N-M+L-1}\gamma_l\right]}\;\;
\end{equation}

\begin{equation}
\times\;\frac{\triangle(\epsilon_1,\ldots
,\epsilon_M;\mu_{L-M+1},\ldots
,\mu_K)}{\triangle^2(\hat\epsilon)\triangle^2(\mu_{L-M+1},\ldots
,\mu_K)\triangle(\mu_1,\ldots ,\mu_{L-M})}\nonumber
\end{equation}

\begin{equation}
\times\;\mbox{det} \left|
\begin{array}{ccc}
  W_{II,N-M+L}(\epsilon_1,\mu_{L-M+1}) & \ldots & W_{II,N-M+L}(\epsilon_1,\mu_{K}) \\
  \vdots &  &  \\
  W_{II,N-M+L}(\epsilon_M,\mu_{L-M+1}) & \ldots & W_{II,N-M+L}(\epsilon_M,\mu_{K}) \\
  W_{I,N-M+L}(\mu_1,\mu_{L-M+1}) & \ldots & W_{I,N-M+L}(\mu_1,\mu_{K}) \\
  \vdots &  &  \\
  W_{I,N-M+L}(\mu_{L-M},\mu_{L-M+1}) & \ldots & W_{I,N-M+L}(\mu_{L-M},\mu_{K})
\end{array}
\right|\nonumber
\end{equation}
where $2L=K+M$ (i.e. the total number of characteristic
polynomials is even) and the kernel functions are defined by
equations (\ref{KernelWI}) and (\ref{WII}).
\end{prop}
\begin{flushleft}\end{flushleft}
The proof of the proposition above can be given by the same method
as for the correlation functions $\F_{II}^K(\hat\epsilon,\hat\mu)$
and $\F_{III}^K(\hat\varpi,\hat\omega)$.
\begin{flushleft}\end{flushleft}
\emph{4.5 Correlation function $\F_{V}^K(\hat\epsilon,
\hat\mu)=\left\langle \frac{\prod\limits_{1}^{K}
\Z_N\left[\mu_l,H\right]}{\prod\limits_1^M\Z_N\left[\epsilon_j,H\right]}
\right\rangle_{H},\;\;K<M$}
\begin{flushleft}\end{flushleft}\nopagebreak[3]
 We also have found the representation
for the correlation function $\F_{V}^K(\hat\epsilon, \hat\mu)$ in
terms of the kernels. Similar to the correlation function
$\F_{IV}^K(\hat\epsilon, \hat\mu)$ the correlation function
$\F_{V}^K(\hat\epsilon, \hat\mu)$  is determined by two kernels.
Both these kernels now  include Cauchy transforms of the orthogonal
polynomials.
 Here
we present the formula without a proof.
\begin{flushleft}\end{flushleft}
\begin{prop}
Let $\Im m\;\epsilon_j\neq 0$. Then the following formula holds
\begin{equation}
\F_{V}^K(\hat\epsilon, \hat\mu)=(-)^{\frac{M(M-1)}{2}}
\;\frac{\left[\gamma_{N-M+L-1}^L\right]}{\left[\prod_{N-M+L}^{N-1}\gamma_s\right]}
\;\frac{1}{\triangle^2(\mu)} \;\frac{1}{M!}\;\;\nonumber
\end{equation}

\begin{equation}
\times\;\sum\limits_{\pi\;\in\;{\textsf{S}}_M}
\frac{\triangle\left(\mu_1,\ldots
,\mu_K;\epsilon_{\pi(1+\frac{M-K}{2})},\ldots
,\epsilon_{\pi(M)}\right)}{\triangle^2\left(\epsilon_{\pi(1+\frac{M-K}{2})},\ldots
,\epsilon_{\pi(M)}\right)\triangle\left(\epsilon_{\pi(1)},\ldots
,\epsilon_{\pi(\frac{M-K}{2})}\right)}\;\;
\end{equation}

\begin{equation}
\times\;\mbox{det}\left|
\begin{array}{ccc}
  W_{II,N-M+L}(\epsilon_{\pi(1+\frac{M-K}{2})},\mu_1) & \ldots & W_{II,N-M+L}(\epsilon_{\pi(M)},\mu_1) \\
  \vdots &  &  \\
 W_{II,N-M+L}(\epsilon_{\pi(1+\frac{M-K}{2})},\mu_K) & \ldots & W_{II,N-M+L}(\epsilon_{\pi(M)},\mu_K) \\
 W_{III,N-M+L}(\epsilon_{\pi(1)},\epsilon_{\pi(1+\frac{M-K}{2})}) & \ldots & W_{III,N-M+L}(\epsilon_{\pi(1)},\epsilon_{\pi(M)}) \\
  \vdots &  &  \\
 W_{III,N-M+L}(\epsilon_{\pi(\frac{M-K}{2})},\epsilon_{\pi(1+\frac{M-K}{2})}) & \ldots & W_{III,N-M+L}(\epsilon_{\pi(\frac{M-K}{2})},\epsilon_{\pi(M)})
\end{array}
\right|\nonumber
\end{equation}
where $2L=K+M$ (i.e. the total number of the characteristic
polynomials is even) and the kernel functions are defined by
equations (\ref{WII}) and (\ref{WIII})\end{prop}
\begin{flushleft}\end{flushleft}
\emph{4.6 Formula for the general correlation function}
\begin{flushleft}\end{flushleft}\nopagebreak[3]
It is possible to derive an exact expression which is valid for
all the five cases considered above. Namely, in \cite{strahov1} we
have proved that
\begin{equation}
\left\langle \frac{
\prod\limits_{j=1}^{K}{\mathcal{Z}}_N\left[\mu_j,
H\right]}{\prod\limits_{j=1}^{M}{\mathcal{Z}}_N\left[\epsilon_j,H\right]}
\right\rangle_{H}=\frac{\prod_{j=N-M}^{N-1}
\gamma_j}{\triangle(\hat\mu)\triangle(\hat \epsilon)}\;\;\nonumber
\end{equation}

\begin{equation}\label{GeneralFormula}
\qquad\qquad\times\;\mbox{det} \left|
\begin{array}{cccc}
  h_{N-M}(\epsilon_1) & h_{N-M+1}(\epsilon_1)
   & \ldots & h_{N+K-1}(\epsilon_1) \\
  \vdots &  &  &  \\
  h_{N-M}(\epsilon_M) & h_{N-M+1}(\epsilon_M)
   & \ldots & h_{N+K-1}(\epsilon_M) \\
  \pi_{N-M}(\mu_1) & \pi_{N-M+1}(\mu_1) & \ldots
   & \pi_{N+K-1}(\mu_1) \\
  \vdots &  &  &  \\
  \pi_{N-M}(\mu_K) & \pi_{N-M+1}(\mu_K)
   & \ldots & \pi_{N+K-1}(\mu_K)
\end{array}\right|
\end{equation}
Here $K$ and $M$ are arbitrary positive integers. Thus formula
(\ref{GeneralFormula}) is also valid when the total number of
characteristic polynomial is odd. However, the new formulae
obtained in this section reveal a kernel structure which makes them more
convenient for investigations of the asymptotic behaviour.
\begin{flushleft}\end{flushleft}
\section{\textbf{Correlation Functions of Characteristic Polynomials and
the Riemann-Hilbert Problem}}
\begin{flushleft}\end{flushleft}\nopagebreak[3]
In this section we establish a
relation between the correlation functions of characteristic
polynomials and the Riemann-Hilbert problem. This relation is
crucial as it enables us to study the asymptotic behaviour
 for the non-Gaussian
case.  Below we review some aspects of the
steepest-descent/stationary phase method for Riemann-Hilbert
problems introduced by Deift and Zhou (a detailed presentation can
be found in the book by Deift \cite{deift1}).

The method is then applied in section 6 for extracting the asymptotics of
the kernels $W_{I,N+K}(\lambda,\mu)$, $W_{II,N}(\epsilon,\mu)$,
and $W_{III,N+K}(\epsilon,\omega)$. These results combined with
the propositions proved in the previous section give us access
to the asymptotic values for all five correlation functions of characteristic
polynomials (\ref{AsymptoticFI})-(\ref{asymptoticFV}).
\begin{flushleft}\end{flushleft}
\emph{5.1 Relation to the Riemann-Hilbert problem.}
\begin{flushleft}\end{flushleft}\nopagebreak[3]
A technique of integrable systems (called the Riemann-Hilbert
problem technique) is exploited in a large number of problems in
mathematics and mathematical physics (see, for example,
\cite{fokaszaharov}). The works of Fokas, Its and Kitaev
\cite{fokas1,fokas2} relate orthogonal polynomials and their
Cauchy transforms
 with an appropriate $2 \times 2$ matrix
Riemann-Hilbert problem thus opening a possibility to apply the
Riemann-Hilbert techniques to orthogonal polynomials and to the
theory of random matrices. The observation of  Fokas, Its and
Kitaev (combined with semi-classical methods for the analysis of
Riemann-Hilbert problems \cite{bleher},
\cite{deift1}-\cite{deift6}) enabled one to understand the
semi-classical asymptotics of the orthogonal polynomials and to
provide an elegant proof of
the Dyson universality conjecture for the Random Matrix Theory
\cite{deift4}-\cite{deift6}, \cite{bleher}.

The exact formulas obtained in the previous sections show that
only monic orthogonal polynomials and their Cauchy transforms
enter various correlation functions of characteristic polynomials.
For this reason the Riemann-Hilbert problem technique arise quite
naturally in the study of correlation functions of that type.

All asymptotic questions we are going to address
turn out to be tractable in the framework of the
Riemann-Hilbert problem for orthogonal polynomials (Fokas, Its and
Kitaev \cite{fokas1,fokas2}). Assume that the contour
$\sum=\mathbb{R}$ is oriented from left to right. The upper side
of the complex plane with respect to the contour will be called
the positive side and the lower part - the negative side. Once the
integer $n\geq 0$ is fixed, the Riemann-Hilbert problem is to find
a $2\times 2$ matrix valued function $Y=Y^{(n)}(z)$ such that the
following conditions are satisfied
\begin{flushleft}\end{flushleft}
\begin{flushleft}\end{flushleft}
\begin{itemize}
  \item $Y^{(n)}(z)- \mbox{analitic}\;\mbox{in}\;
   \mathbb{C}/\mathbb{R}$
  \item $Y^{(n)}_{+}(z)=Y^{(n)}_{-}(z)\left(
  \begin{array}{cc}
    1 & e^{-nV(z)} \\
    0 & 1 \
  \end{array}
  \right),\;z\in \mathbb{R}$
  \item $Y^{(n)}(z)\mapsto\left(I+\mathcal{O}(z^{-1})\right)
  \left(\begin{array}{cc}
    z^n & 0 \\
    0 & z^{-n} \
  \end{array}
  \right)\;\; \mbox{as}\;\; z\mapsto \infty $
\end{itemize}
Here  $Y^{(n)}_{\pm}(z)$ denotes the limit of  $Y^{(n)}(z')$ as
$z'\mapsto z\in \mathbb{R}$ from the positive/negative side. It
was proved by Fokas, Its and Kitaev \cite{fokas1,fokas2} that the
solution of this Riemann-Hilbert problem is unique and is
expressed as
\begin{equation}\label{R-H Solution}
Y^{(n)}(z)=\left(
\begin{array}{cc}
  \pi_n(z) & h_n(z) \\
  \gamma_{n-1}\pi_{n-1}(z) & \gamma_{n-1}h_{n-1}(z)
\end{array}
\right),\;\;\;z\in \mathbb{C}/\mathbb{R}
\end{equation}
where $\pi_n(z)$ and $h_n(z)$ are monic orthogonal polynomials and
their Cauchy transforms, respectively, and the constant $\gamma_{n-1}$ is
defined by equation (\ref{gamma}).

From the propositions of the previous section we immediately
observe that \textit{the correlation functions of characteristic
polynomials are expressible in terms of the elements of
$Y^{(n)}(z)$}. Moreover, the  constant $\gamma_{N-1}$ which
determines pre-factors in the exact expressions for the
correlation functions
(see the propositions (4.1)-(4.5) and the formula (\ref{GeneralFormula}))
emerges in the solution of the Riemann-Hilbert problem as well.

To demonstrate  the relation between the correlation functions of
characteristic polynomials and the Riemann-Hilbert problem
proposed by Fokas, Its and Kitaev it is instructive to consider
the following example. Let us
define the function
\begin{equation}
\F(z)=\left\langle\frac{\mbox{det}(\mu-H)}{\mbox{det}(\epsilon-H)}
\right\rangle_{H,\epsilon=\mu=z},\;\;\;z\in \mathbb{C}
\end{equation}
The function $\F(z)$ is analytic in the whole complex plane, and
$\F(z)\equiv 1,\; \forall z\in \mathbb{C}$. On one hand,
proposition (\ref{FII}) implies that
\begin{equation}
\F(z)=\mbox{det} \left|
\begin{array}{cc}
  \pi_{N}(z) & h_{N}(z) \\
  \gamma_{N-1}\pi_{N-1}(z) & \gamma_{N-1}h_{N-1}(z)
\end{array}
\right|
\end{equation}

\begin{equation}
\;\;\qquad=\mbox{det}[Y^{(N)}(z)]=1,\;\;\; \forall z\in
\mathbb{C}\nonumber
\end{equation}
where $Y^{(N)}(z)$ solves the Riemann-Hilbert problem for
the orthogonal polynomials defined above. On the other  hand, the fact
that $\mbox{det}[Y^{(N)}(z)]=1,\;\;\; \forall z\in \mathbb{C}$ can
be directly obtained  from the definition of the Riemann-Hilbert
problem (Bleher and Its \cite{bleher}). Indeed, if $Y^{(N)}(z)$
solves the Riemann-Hilbert problem for the orthogonal polynomials it
should be that
\begin{equation}
\left[\mbox{det}\;Y^{(N)}\right]_+(z)=\left[\mbox{det}\;Y^{(N)}\right]_-(z)
\;\mbox{det}\left(
\begin{array}{cc}
  1 & e^{-NV(z)} \\
  0 & 1
\end{array}
\right)
\end{equation}

\begin{equation}
\;\qquad\qquad\qquad\quad=\left[\mbox{det}\;Y^{(N)}\right]_-(z)\nonumber
\end{equation}
Therefore, $\mbox{det}\left[ Y^{(N)}(z)\right]$ is analytic in
$\mathbb{C}$ and $\mbox{det}\left[
Y^{(N)}(z)\right]=1+{\mathcal{O}}(z^{-1})$ as
$z\rightarrow\infty$. Then we have $\mbox{det}
\left[Y^{(N)}(z)\right]\equiv 1$.
\begin{flushleft}\end{flushleft}
\emph{5.2 The Deift-Zhou deformations of Riemann-Hilbert
problems.}
\begin{flushleft}\end{flushleft}\nopagebreak[3]
A key ingredient of the Deift-Zhou approach to extracting the asymptotics of
the Riemann-Hilbert problem is the notion of the equilibrium measure (see
\cite{deift1,deift5,deift6}). The equilibrium measure is the
solution of the following energy minimization problem. Assume that
the value $E^{V}$ is defined by
\begin{equation}
E^V=\int V(s)d\mu(s)+\int\int\log|s-t|^{-1}d\mu(s)d\mu(t)
\end{equation}
Then the energy minimization problem is to find a measure
$d\mu(s)$ which minimizes $E^{V}$. On the real line the
equilibrium measure $d\mu(x)=\psi(x)dx$ can be uniquely determined
and  satisfies the following Euler-Lagrange variational conditions:

There exists a real constant $l$ such that
\begin{equation}\label{EulerLagrange1}
2\int\log|x-y|^{-1}d\mu(y)+V(x)\geq l,\;\;\forall x\in \mathbb{R}
\end{equation}
\begin{equation}\label{EulerLagrange2}
2\int\log|x-y|^{-1}d\mu(y)+V(x)=l,\;\;\psi(x)>0
\end{equation}
For the potential $V(x)=x^{2m}$ the solution of the energy
minimization problem described above is given by the equations:
\begin{equation}
d\mu(x)=\psi(x)dx,\;\;\psi(x)=
\frac{m}{i\pi}(x^2-a^2)_+^{1/2}h_1(x)\chi_{(-a,a)}
\end{equation}
\begin{equation}
h_1(x)=x^{2m-2}+
\sum\limits_{j=1}^{m-1}x^{2m-2-2j}a^{2j}\prod\limits_{l=1}^{j}
\frac{2l-1}{2l},\;\;
\end{equation}

\begin{equation}
a=\left(m\prod\limits_{l=1}^{m}\frac{2l-
1}{2l}\right)^{-1/2}\nonumber
\end{equation}
where $\chi_{(a,-a)}$ is the characteristic function of the
interval $(-a,a)$.

When the equilibrium measure is found we can define the following
function :
\begin{equation}
g(z)=\int\log(z-s)\psi(s)ds,\;\;s\in \mathbb{R}, \Im m\;z\neq 0
\end{equation}
Here we take the principal branch of the logarithm, i.e.
\begin{equation}
\log(z-s)=\log|z-s|+i\;\mbox{arg}(z-s)
\end{equation}
where
\begin{equation}
0<\mbox{arg}(z-s)<\pi,\;\;s\in \mathbb{R},\;\;\Im m\;z>0
\end{equation}
\begin{equation}
-\pi<\mbox{arg}(z-s)<0,\;\;s\in \mathbb{R},\;\;\Im m\;z<0
\end{equation}
The function $g(z)$ has the following analytical properties
\begin{itemize}
  \item $g(z)\; \mbox{is}\;\mbox{analytic}\;\mbox{in}\;\mathbb{C}\setminus
  (-\infty,a]$
  \item $g_{\pm}(z)=\int\log|z-s|\psi(s)ds\pm i\pi,\;\;z<-a$
  \item $g_{\pm}(z)=\int\log|z-s|\psi(s)ds\pm
  i\pi\int\limits_{z}^{a}\psi(s)ds,\;\;-a<z<a$
  \item $g(z)=\log z+\mathcal{O}(z^{-
  1}),\;\;z\rightarrow\infty,\;z\in\mathbb{C}\setminus(-\infty,a]$
\end{itemize}

Once the function $g(z)$ and the constant $l$ are specified, the
following transformation is introduced:
\begin{equation}
Y^{(N)}(z)=e^{-\frac{Nl}{2}\sigma_3}m^{(1)}(z)e^{\frac{Nl}{2}\sigma_3}
e^{Ng(z)\sigma_3}\nonumber
\end{equation}
\begin{equation}\label{YM1transform}
\qquad\qquad=\left(
\begin{array}{cc}
  m^{(1)}_{11}(z)e^{Ng(z)} & m^{(1)}_{12}(z)e^{-Nl-Ng(z)} \\
  m^{(1)}_{21}(z)e^{Nl+Ng(z)} &  m^{(1)}_{22}(z)e^{-Ng(z)}
\end{array}
\right)
\end{equation}

 We note that $e^{Ng(z)}$ is analytic in
$\mathbb{C}\setminus[-a,a]$ and
$e^{Ng(z)}=z^{N}(1+\mathcal{O}(z^{-1}))$ as $z\rightarrow\infty$.
It then follows that the matrix function $m^{(1)}(z)$ satisfies
the conditions
\begin{itemize}
  \item
  $m^{(1)}(z)-\mbox{analytic}\;\mbox{in}\;\mathbb{C}/\mathbb{R}$
  \item
  $
  m_{+}^{(1)}(z)=m_{-}^{(1)}(z)
  \left(
  \begin{array}{cc}
    e^{N(g_{-}(z)-g_{+}(z))} & e^{N(g_{+}(z)+g_{-}(z)+l-V(z))} \\
    0 & e^{N(g_{+}(z)-g_{-}(z))}
  \end{array}
  \right)\nonumber
  $
  \item
  $m^{(1)}(z)=I+\mathcal{O}(z^{-1})\;\;\mbox{as}\;\;z\rightarrow\infty
  $
\end{itemize}
 Now the analytical properties of the function
 $g(z)$, the Euler-Lagrange variational conditions, and the explicit
 form of the function $\psi(s)$ can be exploited altogether to derive the
 following representation for the jump matrix $v^{(1)}(z)$ of the
 Riemann-Hilbert problem above:
 \begin{equation}\label{FunctionVarphi(z)}
 v^{(1)}(z)=
  \begin{cases}
  \left(
    \begin{array}{cc}
      1 & e^{-2mN\int\limits_{-a}^{z}(t^2-a^2)^{1/2}h_1(t)dt} \\
      0 & 1 \\
\end{array}
\right), & z\leq -a \\
\left(
    \begin{array}{cc}
      e^{-2mN\int\limits_{z}^{a}(t^2-a^2)^{1/2}_{+}h_1(t)dt} & 1 \\
      0 & e^{2mN\int\limits_{z}^{a}(t^2-a^2)^{1/2}_{+}h_1(t)dt} \
    \end{array}
\right),
     & z\in\;[-a,a]\\
     \left(
    \begin{array}{cc}
      1 & e^{-2mN\int\limits_{-a}^{z}(t^2-a^2)^{1/2}h_1(t)dt} \\
      0 & 1 \\
\end{array}
\right), & z\geq a
  \end{cases}\nonumber
\end{equation}
(In the above formula $(t^2-a^2)^{1/2}$ has positive (negative)
value when $z\geq a$ ($z\leq a$).) Let us set
\begin{equation}
\varphi(z)= m\int\limits_{a}^{z}(t^2-a^2)^{1/2}h_1(t)dt,
\;\;z\in\mathbb{C}\setminus[-a,a]
\end{equation}
The function $\varphi(z)$ is not well-defined as it depends on the
path of integration, but $e^{\pm N\varphi(z)}$ is well-defined and
analytic in $\mathbb{C}\setminus [-a,a]$. With help of the function
$\varphi(z)$ we obtain the following factorization of the jump
matrix $v^{(1)}(z)$:
\begin{equation}
 v^{(1)}(z) = \begin{cases}
  \left(
    \begin{array}{cc}
      1 & e^{-2N\varphi(z)} \\
      0 & 1 \\
\end{array}
\right), & z\leq -a \\
\left(
    \begin{array}{cc}
      1 & 0 \\
      e^{2N\varphi_{-}(z)} & 1
    \end{array}
\right)\left(
    \begin{array}{cc}
      0 & 1 \\
      -1 & 0
    \end{array}
\right)\left(
    \begin{array}{cc}
      1 & 0 \\
      e^{2N\varphi_{+}(z)} & 1
    \end{array}
\right),
     & z\in\;[-a,a]\\
     \left(
    \begin{array}{cc}
      1 & e^{-2mN\int\limits_{-a}^{z}(t^2-a^2)^{1/2}h_1(t)dt} \\
      0 & 1 \\
\end{array}
\right), & z\geq a
  \end{cases}\nonumber
\end{equation}
Once the jump matrix $v^{(1)}(z)$ is factorized we can define the
new matrix-valued function $m^{(2)}(z)$ as  shown in Figure 1.
\begin{flushleft}\end{flushleft}
\begin{picture}(300,250) (-160,-100)
\put (50,50){\circle {100}} \put (50,50) {\line(1,0) {50}} \put
(-40,50) {\line(1,0) {90}} \put (50,50) {\line(1,0) {90}} \put
(20,41) {-$a$} \put (72,41) {$a$}
\put (95,40) {$\mathbb{R}$}
\put (50,50){\vector(1,0){5}} \put (50,70){\vector(1,0){3}} \put
(50,30){\vector(1,0){3}} \put (20,50){\vector(1,0){3}} \put
(80,50){\vector(1,0){3}} \put (100,70) {$m^{(2)}(z)=m^{(1)}(z)$}
\put (100,20) {$m^{(2)}(z)=m^{(1)}(z)$} \put (50,60)
{\line(-5,2){70}} \put (-150,100) {$m^{(2)}(z)=m^{(1)}(z)\left(
\begin{array}{cc}
  1 & 0 \\
  -e^{2N\varphi_{+}(z)} & 1
\end{array}
\right)$} \put (50,40) {\line(-5,-2){70}} \put (-150,0)
{$m^{(2)}(z)=m^{(1)}(z)\left(
\begin{array}{cc}
  1 & 0 \\
  e^{2N\varphi_{-}(z)} & 1
\end{array}
\right)$} \put (-150,-40) {\textbf{Fig}. 1 \begin{small}
Definition of $m^{(2)}(z)$ \end{small} }
\end{picture}
\begin{flushleft}\end{flushleft}
\begin{picture}(300,250) (-110,-100)
\put (50,50){\circle {100}} \put (50,50) {\line(1,0) {50}} \put
(-40,50) {\line(1,0) {90}} \put (50,50) {\line(1,0) {90}} \put
(20,41) {-$a$} \put (72,41) {$a$}
\put (135,40) {$\Sigma_2$}
\put (50,50){\vector(1,0){5}} \put (50,70){\vector(1,0){3}} \put
(50,30){\vector(1,0){3}} \put (20,50){\vector(1,0){3}} \put
(80,50){\vector(1,0){3}} \put (40,50) {\line(-1,-2) {25}} \put
(-20,-10) {\begin{small}$v^{(2)}(z)= \left(
\begin{array}{cc}
  0 & 1 \\
  -1 & 0
\end{array}
\right) $ \end{small}} \put (100,50) {\line(2,-3) {25}} \put
(120,-4) {\begin{small}$v^{(2)}(z)= \left(
\begin{array}{cc}
  1 & e^{-2N\varphi(z)} \\
  0 & 1
\end{array}
\right) $ \end{small}} \put (68,40) {\line(2,-3) {45}} \put
(100,-40) {\begin{small}$v^{(2)}(z)= \left(
\begin{array}{cc}
  1 & e^{2N\varphi(z)} \\
  0 & 1
\end{array}
\right) $ \end{small}} \put (38,66) {\line(-2,3) {13}} \put (0,90)
{\begin{small}$v^{(2)}(z)= \left(
\begin{array}{cc}
  1 & 0 \\
  e^{2N\varphi(z)} & 1
\end{array}
\right) $ \end{small}} \put (-30,50) {\line(-1,-2) {5}} \put
(-95,20) {\begin{small}$v^{(2)}(z)= \left(
\begin{array}{cc}
  1 & e^{-2N\varphi(z)} \\
  0 & 1
\end{array}
\right) $ \end{small}} \put (-110,-70) {\textbf{Fig}. 2
\begin{small} The R-H problem for $m^{(2)}(z)$ \end{small} }
\end{picture}
\begin{flushleft}\end{flushleft}
The matrix $m^{(2)}(z)$ is the solution of the Riemann-Hilbert
problem on the extended contour $\Sigma_2$ (see Fig.2):
\begin{itemize}
  \item $m^{(2)}(z)$ is analytic in $\mathbb{C}/\Sigma_2$
  \item $m_{+}^{(2)}(z)=m_{-}^{(2)}(z)v^{(2)}(z)$
  \item $m^{(2)}(z)\rightarrow I$ as $z\rightarrow \infty$
\end{itemize}
The solution $m^{(2)}(z)$ of the Riemann-Hilbert problem defined
above has the following property:
\begin{prop}\label{technicalproposition}
Let $x\in (-a,a)$. Then for $z$ in the vicinity of $x$ the solution of
the Riemann-Hilbert problem $m^{(2)}(z)$ and its derivative
$dm^{(2)}(z)/dz$ are bounded as $N\rightarrow\infty$.
\end{prop}
\begin{flushleft}\end{flushleft}
Once the proposition (\ref{technicalproposition}) is proved one
can observe that $K_{N}(x,x)$ defined by the equation
(\ref{standardkernel}) is equal to $N\psi(x)$ to the leading
order, i.e.
\begin{equation}\label{KandPsi}
K_{N}(x,x)=N\psi(x)+{\mathcal{O}}(1)\;\;\mbox{as}\;\;N\rightarrow\infty
\end{equation}
\begin{flushleft}\end{flushleft}
\section{\textbf{Asymptotics of the Kernels}}
\begin{flushleft}\end{flushleft}\nopagebreak[3]
In this section we use the results outlined above to determine the
asymptotic behaviour of three kernels $W_{I,N+K}(\lambda,\mu)$,
$W_{II,N}(\lambda,\mu)$, and $W_{III,N-K}(\lambda,\mu)$ in the Dyson's
limit. This is achieved by three subsequent transformations. The
first step is to express the kernels
$W_{I,N+K}(\lambda,\mu)-W_{III,N-K}(\lambda,\mu)$ in terms of
matrix elements of $Y^{(N)}(z)$. We then rewrite them in terms of
$m^{(1)}(z)$ and, finally, in terms of $m^{(2)}(z)$ defined by
Fig. 1 and Fig. 2. The reason for these transformations is that
$m^{(2)}(z)$ and its derivative are bounded matrix valued
functions as $N\rightarrow\infty$. It is this fact and the
equation $\mbox{det}\; m^{(2)}(z)=1$ that enable us to find the large
$N$ asymptotic of kernels in the Dyson's scaling limit. The obtained
asymptotic formulae are summarized in
 Table 3.
\begin{table}[h]
\begin{tabular}{|c|c|}
 \hline
 & \\
  Kernel & Large $N$ asymptotic\\
  & \\
  \hline
   & \\
$W_{I,N+K}(x,\zeta,\eta)$ &  $
  c^2_N\left[N\rho(x)\right]e^{NV(x)}e^{\alpha(x)(\zeta+\eta)}\mathbb{S}_I(\zeta-\eta)$ \\
  & \\
  \hline
   & \\
  $W_{II,N}(x,\zeta,\eta)$, & $
  -\frac{c^2_N}{2\pi i}\left[N\rho(x)\right]\;e^{-\alpha(x)(\zeta-\eta)}\mathbb{S}_{II}(\zeta-\eta)$\\
   & \\
   \hline
   & \\
  $W_{III,N-K}(x,\zeta,\eta)$ & $
  -\frac{c^2_N}{2\pi i}\left[N\rho(x)\right]\;e^{-NV(x)}
  \;e^{-\alpha(x)(\zeta+\eta)}\mathbb{S}_{III}(\zeta-\eta)$ \\
 & \\
\hline
\end{tabular}
\\
\small{\caption{Asymptotic of kernels}}
  \end{table}
  \begin{flushleft}\end{flushleft}
\emph{6.1 Large $N$ limit of\;\; $W_{I,N+K}(\lambda,\mu)$}
\begin{flushleft}\end{flushleft}
 We fix a point $x\in (-a,a)$. The
interval $(-a, a)$ is the support of the equilibrium measure for
the given potential function $V(x)$. For simplicity we  assume
that the support of the equilibrium measure includes only one
interval. This is the case for the potential function
$V(x)=x^{2m}$, $m\geq 1$. Introduce new coordinates $\zeta,\eta$
\begin{equation}\label{lambdax}
\lambda=x+\frac{\zeta}{N\rho(x)},\;\;\mu=x+\frac{\eta}{N\rho(x)}
\end{equation}
In what follows we  consider the Dyson's scaling limit. In such a
 limit the difference between points $\lambda$ and $\mu$
goes to zero, the size $N$ goes to infinity, the product
$N(\lambda-\mu)$ remains finite. In equation (\ref{lambdax})
$\rho(x)=K_N(x,x)/N$ is the density of states, with the kernel
$K_N(x,x)$ being given by the equation (\ref{standardkernel}). In
terms of the
new coordinates the kernel $W_{I,N+K}(\lambda,\mu)\equiv
W_{I,N+K}(x,\zeta,\eta)$ is expressed as a determinant of monic
orthogonal polynomials,
\begin{equation}
W_{I,N+K}(x,\zeta,\eta)=\left[N\rho(x)\right]\left[\zeta-\eta\right]^{-1}
\nonumber
\end{equation}

\begin{equation}\label{WPolynomials}
\qquad\qquad\qquad\quad\times\;\mbox{det} \left|
\begin{array}{cc}
  \pi_{N}(x+\frac{\zeta}{N\rho(x)})
   & \pi_{N}(x+\frac{\eta}{N\rho(x)}) \\
  \pi_{N-1}(x+\frac{\zeta}{N\rho(x)})
   & \pi_{N-1}(x+\frac{\eta}{N\rho(x)})
\end{array}
\right|
\end{equation}
Thus the problem about asymptotics of the two-point kernel function
$W_{I,N+K}(\lambda,\mu)$ is reduced to the investigation of the
large $N$ asymptotics of the determinant in the equation
(\ref{WPolynomials}).
\begin{flushleft}\end{flushleft}
\begin{prop}\label{LargeNW}(Large $N$ asymptotics of the kernel
 $W_{I,N+K}(x,\zeta,\eta)$).\\ Let $x\in(-a,a)$,
 $-\theta\leq \zeta,\eta\leq\theta$, $\zeta\neq \eta$
 and $N\gg K$.
 Then in the large $N$ limit the kernel function
 $W_{I,N+K}(x,\zeta,\eta)$ is related with the universal
 kernel
 $S_{I}(\zeta-\eta)$
 as
\begin{equation}
W_{I,N+K}(x,\zeta,\eta)=\left[c_{N}\right]^{2K}\;\left[N\rho(x)\right]\;e^{NV(x)}
e^{\alpha(x)(\zeta+\eta)}\left[S_{I}(\zeta-\eta)+
{\mathcal{O}}(1/N)\right]
\end{equation}
where the universal kernel $\mathbb{S}_{I}(\zeta-\eta)$ is
\begin{equation}
S_{I}(\zeta-\eta)=\frac{\sin\left(\pi(\zeta-\eta)\right)}{\pi(\zeta-\eta)}
\end{equation}
\end{prop}
\begin{flushleft}\end{flushleft}
\begin{proof}
 Equation (\ref{R-H Solution}) enables us to write the kernel
$W_{I,N+K}(x,\zeta,\eta)$ in the form:
\begin{equation}
W_{I,N+K}(x,\zeta,\eta)=
\left[\gamma_{N+K-1}\right]^{-1}\left[N\rho(x)\right]\left[\zeta-\eta\right]^{
-1}\;\nonumber
\end{equation}

\begin{equation}\label{WYDeterminant}
\qquad\qquad\times\;\mbox{det} \left|
\begin{array}{cc}
  Y_{11}^{(N+K)}(x+\frac{\zeta}{N\rho(x)})
   & Y_{11}^{(N+K)}(x+\frac{\eta}{N\rho(x)}) \\
  Y_{21}^{(N+K)}(x+\frac{\zeta}{N\rho(x)})
   & Y_{21}^{(N+K)}(x+\frac{\eta}{N\rho(x)})
\end{array}
\right|
\end{equation}
The large $N$ asymptotics is completely determined by the
determinant\footnote{Since $N\gg K$, we do not distinguish
between $N$ and $N+K$ when studying the asymptotics.} in
the equation above. Large $N$ limit of this determinant is
considered in Deift's book \cite{deift1}, Chapter 8. Here we
reproduce his derivation. Equation (\ref{YM1transform}) gives
\begin{equation}
\mbox{det} \left|
\begin{array}{cc}
  Y_{11}^{(N+K)}(x+\frac{\zeta}{N\rho(x)})
   & Y_{11}^{(N+K)}(x+\frac{\eta}{N\rho(x)}) \\
  Y_{21}^{(N+K)}(x+\frac{\zeta}{N\rho(x)})
   & Y_{21}^{(N+K)}(x+\frac{\eta}{N\rho(x)})
\end{array}\nonumber
\right|
\end{equation}

\begin{equation}
=e^{N\left[g_+(x+\zeta/N\rho(x))+g_+(x+\eta/N\rho(x))+l\right]}
\end{equation}

\begin{equation}
\times\;\mbox{det} \left|
\begin{array}{cc}
  m_{11}^{(1)}(x+\frac{\zeta}{N\rho(x)})
   & m_{11}^{(1)}(x+\frac{\eta}{N\rho(x)}) \\
  m_{21}^{(1)}(x+\frac{\zeta}{N\rho(x)})
   & m_{21}^{(1)}(x+\frac{\eta}{N\rho(x)})
\end{array}\nonumber
\right|
\end{equation}
where $m^{(1)}(z)$ is the solution of the transformed
Riemann-Hilbert problem (see the previous section). Rewrite the
determinant in terms of $m^{(2)}(z)$ (see Fig. 1 and Fig. 2 where
$m^{(2)}(z)$ is related with $m^{(1)}(z)$). This leads to the
following expression for the kernel
\begin{equation}
W_{I,N+K}(x,\zeta,\eta)=
\left[\gamma_{N+K-1}\right]^{-1}\left[N\rho(x)\right]\left[\zeta-\eta\right]^{
-1}\;\nonumber
\end{equation}

\begin{equation}
\times\;e^{N\left[g_+(x+\zeta/N\rho(x))+g_+(x+\eta/N\rho(x))+l\right]}
\end{equation}

\begin{equation}
\times\; \mbox{det}
 \begin{small}\left|
 \begin{array}{cc}
  \left[m_{11}^{(2)}(x_{\zeta})\right]_+
   +\left[m_{12}^{(2)}(x_{\zeta})\right]_+e^{2N\varphi_{+}(x_{\zeta})}
   & \left[m_{11}^{(2)}(x_{\eta})\right]_+
   +\left[m_{12}^{(2)}(x_{\eta})\right]_+e^{2N\varphi_{+}(x_{\eta})}\nonumber \\
\left[m_{21}^{(2)}(x_{\zeta})\right]_+
   +\left[m_{22}^{(2)}(x_{\zeta})\right]_+e^{2N \varphi_{+}(x_{\zeta})}
   & \left[m_{21}^{(2)}(x_{\eta})\right]_+
   +\left[m_{22}^{(2)}(x_{\eta})\right]_+e^{2N
   \varphi_{+}(x_{\eta})}\nonumber
\end{array}
\right |
\end{small}
\end{equation}
where $x_{\zeta}=x+\frac{\zeta}{N\rho(x)}$,
$x_{\eta}=x+\frac{\eta}{N\rho(x)}$. Remember that  $m^{(2)}(z)$ and its
derivative are bounded matrix valued functions in the vicinity of
the point $x$ and $\mbox{det}\;m^{(2)}(z)=1$. This observation
enables one to expand all functions in the determinant in the vicinity
of the point $x$, and to find the leading term:
\begin{equation}
W_{I,N+K}(x,\zeta,\eta)=
\left[\gamma_{N+K-1}\right]^{-1}\left[N\rho(x)\right]\left[\zeta-\eta\right]^{
-1}\;\nonumber
\end{equation}

\begin{equation}
\times\;e^{N\left[g_+(x+\zeta/N\rho(x))+g_+(x+\eta/N\rho(x))+l\right]}
\end{equation}

\begin{equation}
\times\;\left[e^{2N\varphi_+(x+\eta/N\rho(x))}-e^{2N\varphi_+(x+\zeta/N\rho(x))}\right]
\left[1+{\mathcal{O}}(N^{-1})\right] \nonumber
\end{equation}
The next step is to express the functions
$g_{+}(x+\zeta/N\rho(x))$, $g_{+}(x+\eta/N\rho(x))$ in terms of
the functions $\varphi_+(x_{\zeta})$, $\varphi_+(x_{\eta})$. It
can be done as follows. Let $z\in(-a,a)$.
 When $z\in(-a,a)$ the
following equations hold (see properties of the function
$g_{\pm}(z)$ summarized in the previous section):
\begin{equation}
g_+(z)+g_-(z)=2\int\log|z-s|\psi(s)ds
\end{equation}
\begin{equation}
g_+(z)-g_-(z)=2i\pi\int_z^a\psi(s)ds
\end{equation}
We use the Euler-Lagrange variational condition
(\ref{EulerLagrange2}) together with the equations above and find
that the functions $g_{\pm}(z)$ are completely determined by the
function $\varphi_{+}(z)$, by the potential $V(z)$ and the
constant $l$. Namely,
\begin{equation}\label{ConsequenceForG}
\begin{cases}
    g_{+}(z)=-\varphi_{+}(z)+V(z)-\frac{l}{2} &  \\
    g_{-}(z)=\;\;\varphi_{+}(z)+V(z)-\frac{l}{2} &
  \end{cases}
\end{equation}
This gives
\begin{equation}
W_{I,N+K}(x,\zeta,\eta)=
\left[\gamma_{N+K-1}\right]^{-1}\left[N\rho(x)\right]\left[\zeta-\eta\right]^{
-1}\;e^{NV(x)}e^{\alpha(x)(\zeta+\eta)}\nonumber
\end{equation}

\begin{equation}
\times\left[e^{N\left[\varphi_+(x+\zeta/N\rho(x))-\varphi_+(x+{\eta}/N\rho(x))\right]}-
e^{-N\left[\varphi_+(x+\eta/N\rho(x))-\varphi_+(x+\zeta/N\rho(x))\right]}
\right]
\end{equation}
In the vicinity of the point $x$ we have
\begin{equation}
\varphi_+(x+\frac{\zeta}{N\rho(x)})=\varphi_+(x)+i\pi\psi(x)\frac{\zeta}{N\rho(x)}+
{\mathcal{O}}\left(\frac{1}{N^2}\right)
\end{equation}
A similar expression is obtained for
$\varphi_+(x+\frac{\eta}{N\rho(x)})$. We complete the proof using
the equation (\ref{KandPsi}) and the relation
 $K_N(x,x)=N\rho(x)$.
\end{proof}
\begin{flushleft}\end{flushleft}
\emph{6.2 Large $N$ limit of $W_{II,N}(\epsilon,\mu)$}
\begin{flushleft}\end{flushleft}\nopagebreak[3]
In this subsection we investigate Dyson's scaling limit of the
kernel $W_{II,N}(\epsilon,\mu)$  constructed from monic orthogonal
polynomials and their Cauchy transforms. Similar to the procedure
applied to the kernel function $W_{I,N+K}(\epsilon,\mu)$ we
introduce new coordinates,
\begin{equation}
\epsilon=x+\frac{\zeta}{N\rho(x)},\;\;\;\mu=x+\frac{\eta}{N\rho(x)}
\end{equation}
In the coordinates $\zeta, \eta$ the kernel
$W_{II,N}(\epsilon,\mu)\equiv W_{II,N}(x,\zeta,\eta)$ has the
form:
\begin{equation}
W_{II,N}(x,\zeta,\eta)=
\left[N\rho(x)\right]\left[\eta-\zeta\right]^{-1}\nonumber
\end{equation}

\begin{equation}\label{NewKernelAsDetPol}
 \qquad\qquad\qquad\;\times\;\mbox{det} \left|
\begin{array}{cc}
  \pi_{N}(x+\frac{\eta}{N\rho(x)})
   & h_{N}(x+\frac{\zeta}{N\rho(x)}) \\
  \pi_{N-1}(x+\frac{\eta}{N\rho(x)})
   & h_{N-1}(x+\frac{\zeta}{N\rho(x)})
\end{array}
\right|
\end{equation}
We can see that in order to derive an asymptotic expression for
the kernel $W_{II,N}(x,\zeta,\eta)$ we need the asymptotics of the
determinant which contains monic orthogonal polynomials and their
Cauchy transform. The Riemann-Hilbert technique proves to be a convenient
tool in this case as well.
\begin{flushleft}\end{flushleft}
\begin{prop}(Large $N$ asymptotics for the kernel
$W_{II,N}(x,\zeta,\eta)$)\\ Let $x\in (-a,a)$, $-\theta\leq
\zeta,\eta\leq\theta$, $\zeta\neq\eta$, $\Im m\; \zeta\neq 0$.
Then the following asymptotic expression for the kernel
$W_{II,N}(x,\zeta,\eta)$ holds:
\begin{equation}
W_{II,N}(x,\zeta,\eta)=\left[\gamma_{N}\right]^{-1}
\left[N\rho(x)\right]e^{-\alpha(x)(\zeta-\eta)}
\left[S_{II}(\zeta-\eta)+{\mathcal{O}}(1/N)\right]
\end{equation}
The universal two-point kernel $S_{II}(\zeta-\eta)$ is expressed
as
\begin{equation}
S_{II}(\zeta-\eta)=
  \begin{cases}
    \frac{e^{i\pi(\zeta-\eta)}}{\zeta-\eta} & \Im m\;\zeta>0, \\
    \frac{e^{-i\pi(\zeta-\eta)}}{\zeta-\eta} & \Im m\;\zeta<0
  \end{cases}
\end{equation}
\end{prop}
\begin{flushleft}\end{flushleft}
\begin{proof}
Express the determinant in the equation (\ref{NewKernelAsDetPol}) in
terms of $Y^{(N)}(z)$ which is the matrix valued solution of the
Riemann Hilbert problem for the orthogonal polynomials. We have
\begin{equation}
W_{II,N}(x,\zeta,\eta)=
\left[N\rho(x)\right]\left[\eta-\zeta\right]^{-1}\left[\gamma_{N-1}\right]^{-1}\nonumber
\end{equation}

\begin{equation}\label{NewKernelAsDetY}
 \qquad\qquad\qquad\;\times\;\mbox{det} \left|
\begin{array}{cc}
  Y^{(N)}_{11}(x+\frac{\eta}{N\rho(x)})
   & Y^{(N)}_{12}(x+\frac{\zeta}{N\rho(x)}) \\
  Y^{(N)}_{21}(x+\frac{\eta}{N\rho(x)})
   & Y^{(N)}_{22}(x+\frac{\zeta}{N\rho(x)})
\end{array}
\right|
\end{equation}
Equation (\ref{YM1transform}) enables one to replace the elements of
$Y^{(N)}$ by the elements of $m^{(1)}(z)$ (we
remind that $m^{(1)}(z)$ is the solution of the transformed
Riemann-Hilbert problem). Then we obtain
\begin{equation}
W_{II,N}(x,\zeta,\eta)=
\left[N\rho(x)\right]\left[\eta-\zeta\right]^{-1}\left[\gamma_{N-1}\right]^{-1}
e^{N\left[g(x+\eta/N\rho(x))-g(x+\zeta/N\rho(x))\right]}\nonumber
\end{equation}

\begin{equation}\label{NewKernelAsDetM1}
 \qquad\qquad\qquad\;\times\;\mbox{det} \left|
\begin{array}{cc}
  m^{(1)}_{11}(x+\frac{\eta}{N\rho(x)})
   & m^{(1)}_{12}(x+\frac{\zeta}{N\rho(x)}) \\
  m^{(1)}_{21}(x+\frac{\eta}{N\rho(x)})
   & m^{(1)}_{22}(x+\frac{\zeta}{N\rho(x)})
\end{array}
\right|
\end{equation}
The elements of $m^{(1)}(z)$ can be further replaced by the
elements of $m^{(2)}(z)$ (see Fig. 1 and Fig. 2 where the
Riemann-Hilbert problem for $m^{(2)}(z)$ is specified and the
relation between $m^{(1)}(z)$ and $m^{(2)}(z)$ is shown).
Again, $m^{(2)}(z)$ and its derivative are bounded near the point $x$,
and $\mbox{det}\;m^{(2)}(z)=1$. Thus we can expand around the
point $x$ and show that the determinant in equation
(\ref{NewKernelAsDetM1}) is equal to one to the leading order.
Therefore,
\begin{equation}
W_{II,N}(x,\zeta,\eta)=
\left[N\rho(x)\right]\left[\eta-\zeta\right]^{-1}\left[\gamma_{N-1}\right]^{-1}
e^{N\left[g(x+\eta/N\rho(x))-g(x+\zeta/N\rho(x))\right]}\left[1+{\mathcal{O}}(1/N)\right]\nonumber
\end{equation}
Introduce $z_1=x+\eta/N\rho(x)$, $z_2=x+\zeta/N\rho(x)$. Then we
have
\begin{equation}
N\left[g(z_1)-g(z_2)\right]\;\;=N\left[\int\log(z_1-s)\psi(s)ds-\int\log(z_2-s)\psi(s)ds\right]
\nonumber \end{equation}

\begin{equation}
\qquad\qquad\qquad\qquad=N\int\log\left[\frac{z_1-s}{z_2-s}\right]\psi(s)ds
\nonumber
\end{equation}

\begin{equation}
\qquad\qquad\qquad\qquad=N\int\log\left[1+\frac{z_1-z_2}{z_2-s}\right]\psi(s)ds
\end{equation}

\begin{equation}
\qquad\qquad\qquad\qquad=\frac{[\eta-\zeta]}{\rho(x)}\;
\lim\limits_{\delta\rightarrow 0}
\left[\int\limits_{-\infty}^{+\infty}\frac{\psi(s)ds}{x-s\pm
i\delta}\right]+{\mathcal{O}}(1/N)\nonumber
\end{equation}

\begin{equation}
\qquad\qquad\qquad\qquad=\frac{[\eta-\zeta]}{\rho(x)}\;\left[\pi
H\psi(x)\mp i\pi\psi(x)\right]\nonumber
\end{equation}
 where $+$ ($-$) corresponds to the
positive (negative) imaginary part of $\zeta$, and $H\psi(x)$
stands for the Hilbert transform of $\psi(x)$,
\begin{equation}
H\psi(x)=\frac{1}{\pi}\;P.V.\;\int\frac{\psi(s)ds}{x-s}
\end{equation}
Furthermore, from equation (\ref{EulerLagrange2}) (the second
Euler-Lagrange condition) we observe that
\begin{equation}
H\psi(x)=\frac{1}{2\pi}\;V'(x)
\end{equation}
Now use that $\psi(x)$ is equal to the density of states in the large $N$
limit to obtain
\begin{equation}
N\left[g(z_1)-g(z_2)\right]=\pm
i\pi(\zeta-\eta)-\frac{V'(x)}{2\rho(x)}(\zeta-\eta)+{\mathcal{O}}(\frac{1}{N})
\end{equation}
which completes the proof.
\end{proof}
\newpage
\begin{flushleft}\end{flushleft}
\emph{6.3 Large $N$ limit of $W_{III,N-K}(\epsilon,\omega)$}
\begin{flushleft}\end{flushleft}
\nopagebreak[3]
Finally we investigate  the Dyson's scaling limit for the kernel
$W_{III,N-K}(\epsilon,\omega)$. This kernel is constructed from
the Cauchy transforms of monic orthogonal polynomials, and the
transforms are not analytic. For this reason we need to consider
different situations corresponding to different signs
of the imaginary parts. Just as
before  we introduce new coordinates
\begin{equation}
\epsilon=x+\zeta/N\rho(x),\;\;\;\omega=x+\eta/N\rho(x)
\end{equation}
where  $\Im m\;\zeta\neq 0,\;\Im m\;\eta\neq 0$. In terms of these
coordinates we define  $W_{III,N-K}(x,\zeta,\eta)\equiv
W_{III,N-K}(x+\frac{\zeta}{N\rho(x)},x+\frac{\eta}{N\rho(x)})$.
\begin{flushleft}\end{flushleft}
\begin{prop}
(Large $N$ asymptotics of the kernel $W_{III,N-K}(x,\zeta,\eta)$).
Let $x\in(-a,a)$, $-\theta\leq \zeta,\eta\leq\theta$, $\Im
m\;\zeta\neq 0,\;\Im m\;\eta\neq 0$. Then the large $N$ limit of
the kernel function $W_{III,N-K}(x,\zeta,\eta)$ is related to a
universal kernel $S_{III}(\zeta,\eta)$ as
\begin{equation}
W_{III,N-K}(x,\zeta,\eta)\nonumber
\end{equation}
\begin{equation}
\qquad=\left[\gamma_N\right]^{-1}\left[N\rho(x)\right]
e^{-NV(x)}e^{-\alpha(x)(\zeta+\eta)}\left[S_{III}(\zeta,\eta)+{\mathcal{O}}(N^{-1})\right]
\end{equation}
where the universal kernel $S_{III}(\zeta,\eta)$ is given by the
formula
\begin{equation}
S_{III}(\zeta,\eta)=
  \begin{cases}
    \frac{1}{\zeta-\eta} & \Im m\;\zeta >0,\Im m\;\eta<0 , \\
    \frac{1}{\eta-\zeta} & \Im m\;\zeta <0,\Im m\;\eta>0, \\
    0 & \text{otherwise}
  \end{cases}
\end{equation}
\end{prop}
\begin{flushleft}\end{flushleft}
\begin{proof}
We  give the proof only for the case $\Im m\;\zeta
> 0,\;\Im m\;\eta < 0$. The other three cases with different signs
of imaginary parts can be considered in a similar manner. The
kernel $W_{III,N-K}(x,\zeta,\eta)$ can be expressed as
\begin{equation}
W_{III,N-K}(x,\zeta,\eta)=\left[N\rho(x)\right]\left[\zeta-\eta\right]^{-1}\nonumber
\end{equation}

\begin{equation}
\qquad\qquad\qquad\qquad\times\;\mbox{det} \left|
\begin{array}{cc}
  h_{N-K}(x+\zeta/N\rho(x)) & h_{N-K}(x+\eta/N\rho(x)) \\
 h_{N-K-1}(x+\zeta/N\rho(x)) & h_{N-K-1}(x+\eta/N\rho(x))
\end{array}
\right|\nonumber
\end{equation}
Now we exploit the relation to the Riemann-Hilbert problem for
the orthogonal polynomials. We replace the Cauchy transforms in the
determinant above by the corresponding elements of the matrix
$Y^{(N-K)}$ which is the solution of the Riemann-Hilbert problem.
We have
\begin{equation}
W_{III,N-K}(x,\zeta,\eta)=
\left[N\rho(x)\right]\left[\zeta-\eta\right]^{-1}\left[\gamma_{N-K}\right]^{-1}\nonumber
\end{equation}

\begin{equation}
\qquad\qquad\qquad\qquad\times\;\mbox{det} \left|
\begin{array}{cc}
  \left[Y^{(N-K)}_{12}(x+\zeta/N\rho(x))\right]_+ & \left[Y^{(N-K)}_{12}(x+\eta/N\rho(x))\right]_- \\
\left[Y^{(N-K)}_{22}(x+\zeta/N\rho(x))\right]_+ &
\left[Y^{(N-K)}_{22}(x+\eta/N\rho(x))\right]_-
\end{array}
\right|\nonumber
\end{equation}
Then we employ the transformation (equation (\ref{YM1transform})) from the
solution $Y^{(N-K)}(z)$ of the original Riemann-Hilbert problem to
that of the new Riemann-Hilbert problem $m^{(1)}(z)$ defined by
the jump matrix $v^{(1)}(z)$ ($N$ should be replaced by $N-K$).
This yields
\begin{equation}
W_{III,N-K}(x,\zeta,\eta)=\left[N\rho(x)
\right]\left[\zeta-\eta\right]^{-1}\left[\gamma_{N-K}\right]^{-1}
\nonumber
\end{equation}

\begin{equation}
\;\qquad\qquad\qquad\qquad\times\;
\exp\left[-(N-K)\left[l+g_+(x+\zeta/N\rho(x))+g_+(x+\eta/N\rho(x))\right]\right]
\nonumber
\end{equation}

\begin{equation}
 \;\qquad\qquad\qquad\qquad\times\;
\mbox{det} \left|
\begin{array}{cc}
  \left[m^{(1)}_{12}(x+\zeta/N\rho(x))\right]_+ & \left[m^{(1)}_{12}(x+\eta/N\rho(x))\right]_- \\
\left[m^{(1)}_{22}(x+\zeta/N\rho(x))\right]_+ &
\left[m^{(1)}_{22}(x+\eta/N\rho(x))\right]_-
\end{array}
\right|\nonumber
\end{equation}
In turn, the function $m^{(1)}(z)$ is related to the matrix valued
function $m^{(2)}(z)$ (see Fig. 2) which is the solution of the
deformed Riemann-Hilbert problem defined by Fig. 2. Correspondingly,
we rewrite the
kernel in terms of the elements of $m^{(2)}(z)$,
\begin{equation}
W_{III,N-K}(x,\zeta,\eta)=\left[N\rho(x)
\right]\left[\zeta-\eta\right]^{-1}\left[\gamma_{N-K}\right]^{-1}
\nonumber
\end{equation}

\begin{equation}
\;\qquad\qquad\qquad\qquad\times\;
\exp\left[-(N-K)\left[l+g_+(x+\zeta/N\rho(x))+g_+(x
+\eta/N\rho(x))\right]\right]\nonumber
\end{equation}

\begin{equation}
 \;\qquad\qquad\qquad\qquad\times\;
\mbox{det} \left|
\begin{array}{cc}
  \left[m^{(2)}_{12}(x+\zeta/N\rho(x))\right]_+ & \left[m^{(2)}_{12}(x+\eta/N\rho(x))\right]_- \\
\left[m^{(2)}_{22}(x+\zeta/N\rho(x))\right]_+ &
\left[m^{(2)}_{22}(x+\eta/N\rho(x))\right]_-
\end{array}
\right|\nonumber
\end{equation}
As $m^{(2)}(z)$ and $dm^{(2)}(z)/dz$ are bounded matrix valued
functions in the vicinity of the point $x$, and
$g_+(z)+g_+(z)+l=V(z)$ for $z\in (-a,a)$ (see equation
(\ref{ConsequenceForG})) we obtain  the following large $N$ limit
for the kernel $W_{III,N-K}(x,\zeta,\eta)$
\begin{equation}
W_{III,N-K}(x,\zeta,\eta)=\left[N\rho(x)
\right]\left[\zeta-\eta\right]^{-1}\left[\gamma_{N-K}\right]^{-1}
\;e^{-NV(x)}e^{-\alpha(x)(\zeta+\eta)}\nonumber
\end{equation}

\begin{equation}
 \;\qquad\qquad\qquad\qquad\times\;
\left[\mbox{det} \left|
\begin{array}{cc}
  \left[m^{(2)}_{11}(x)\right]_+ & \left[m^{(2)}_{12}(x)\right]_- \\
\left[m^{(2)}_{22}(x)\right]_+ & \left[m^{(2)}_{22}(x)\right]_-
\end{array}
\right|+\mathcal{O}(1/N)\right]\nonumber
\end{equation}
However, in the vicinity of the point $x$ we have (see Fig. 2)
\begin{equation}
\left(
\begin{array}{cc}
  \left[m^{(2)}_{121}(x)\right]_+ & \left[m^{(2)}_{12}(x)\right]_+ \\
\left[m^{(2)}_{21}(x)\right]_+ & \left[m^{(2)}_{22}(x)\right]_+
\end{array}
\right) = \left(
\begin{array}{cc}
  \left[m^{(2)}_{11}(x)\right]_- & \left[m^{(2)}_{12}(x)\right]_- \\
\left[m^{(2)}_{21}(x)\right]_- & \left[m^{(2)}_{22}(x)\right]_-
\end{array}
\right)\; \left(
\begin{array}{cc}
  0 & 1 \\
  -1 & 0
\end{array}
\right)\nonumber
\end{equation}
i.e.
\begin{equation}
\left[m^{(2)}_{12}(x)\right]_-=-\left[m^{(2)}_{11}(x)\right]_+,\;\;
\left[m^{(2)}_{21}(x)\right]_-=-\left[m^{(2)}_{22}(x)\right]_+\nonumber
\end{equation}
Taking into account that
\begin{equation}
\mbox{det} \left|
\begin{array}{cc}
  \left[m^{(2)}_{11}(x)\right]_+ & \left[m^{(2)}_{12}(x)\right]_- \\
\left[m^{(2)}_{21}(x)\right]_+ & \left[m^{(2)}_{22}(x)\right]_-
\end{array}
\right|=1\nonumber
\end{equation}
we finally obtain
\begin{equation}
W_{III,N-K}(x,\zeta,\eta)=\left[N\rho(x)\right]\left[\zeta
-\eta\right]^{-1}\left[\gamma_{N-K}\right]^{-1}
\;e^{-NV(x)}e^{-\alpha(x)(\zeta+\eta)}\;\left[1+{\mathcal{O}}(1/N)\right]\nonumber
\end{equation}
\end{proof}
\begin{flushleft}\end{flushleft}
\section{\textbf{Negative Moments}}
\begin{flushleft}\end{flushleft}
In this section we derive the asymptotic expression
(\ref{NegativeMomentAsymptotic}) for the negative moments of
characteristic polynomials $\M_{x,N}^K(\delta)$ defined by
the equation (\ref{NegativeMomentsDefinition}).  The negative moments
can be obtain as limiting values of the correlation function
$\F_{III}^K(\hat x+\hat \zeta/N\rho(x))$. The large $N$ asymptotic
for that function is given by the equation (\ref{ASYMPTOTICFIII}). We
define three $K$ dimensional vectors, $\hat\zeta^+$,
$\hat\zeta^-$, and $\hat\delta$. The components of the vectors
$\hat\zeta^+$ ($\hat\zeta^-$) are pure imaginary and have positive
(negative) imaginary parts. The vector $\hat\delta$ has all components
equal to each other and equal to the real parameter $\delta$ in
the definition of the negative moments (equation
(\ref{NegativeMomentsDefinition})). We exploit our asymptotic
formula (\ref{ASYMPTOTICFIII}) for the correlation function
$\F_{III}^K(\hat x+\hat \zeta/N\rho(x))$ and write the following
expression for the negative moments:
\begin{equation}\label{NM71}
\M_{x,N}^K(\delta)=
(-)^K\left[\gamma_N\right]^K\left[N\rho(x)\right]^{K^2}
e^{-KNV(x)}\frac{\J_K(\delta)}{(2K)!}
\end{equation}
where
\begin{equation}
\J_K(\delta) = \lim\limits_{\hat\zeta^{\pm}\rightarrow\pm
i\hat\delta/2}\left[
 \sum\limits_{\pi\;\in\; {\textsf{S}}_{2K}}
\frac{\mbox{det}\left[\mathbb{S}_{III}\left(\zeta_{\pi(i)},\zeta_{\pi(j+K)}\right)\right]_{
1\leq i,j\leq K}}{\triangle(\zeta_{\pi(1)},\ldots,\zeta_{\pi(K)})
\triangle(\zeta_{\pi(K+1)},
\ldots,\zeta_{\pi(2K)})}\right]\nonumber
\end{equation}
 Following the method by Brezin and Hikami's
\cite{brezin1,brezin2} we represent the determinant of the kernel
divided by two Vandermonde determinants as a contour integral,
i.e.
\begin{equation}\label{ContourIntegralRepresentation}
\sum\limits_{\pi\;\in\; {\textsf{S}}_{2K}}
\frac{\mbox{det}\left[\mathbb{S}_{III}\left(\zeta_{\pi(i)},\zeta_{\pi(j+K)}\right)\right]_{
1\leq i,j\leq K}}{\triangle(\zeta_{\pi(1)},\ldots,\zeta_{\pi(K)})
\triangle(\zeta_{\pi(K+1)}, \ldots,\zeta_{\pi(2K)})}
\end{equation}

\begin{equation}
=\frac{1}{K!}\sum\limits_{\pi\;\in\; {\textsf{S}}_{2K}} \oint\oint
\prod\limits_1^K\frac{du_idv_i}{(2\pi i)^2} \frac{\triangle(\hat
u) \triangle(\hat
v)\prod_1^K\mathbb{S}_{III}(u_i,v_i)}{\prod\limits_{i,j=1}^K
\left[u_i-\zeta_{\pi(j)}\right]\left[v_i-\zeta_{\pi(K+j)}\right]}\nonumber
\end{equation}
where the contours of integration are chosen in such a way that
all components of the vector $\hat \zeta$ give rise to
contributions to the integral as simple poles. We have to compute
the integral above when
\begin{equation}
\hat\zeta=(i\delta/2,\ldots ,i\delta/2,-i\delta/2,\ldots
,-i\delta/2)\equiv(\hat\zeta^+,\hat\zeta^-)
\end{equation}
All those permutations $\pi\;\in\;{\textsf{S}}_{2K}$ that lead to
the same vector $\pi \hat\zeta$ produce the same contribution to
the integral. This observation permits us to rewrite the right-hand
side of the equation (\ref{ContourIntegralRepresentation}) as
\begin{equation}
K!\;\sum\limits_{\sigma} \oint\oint
\prod\limits_1^K\frac{du_idv_i}{(2\pi i)^2} \frac{\triangle(\hat
u) \triangle(\hat
v)\prod_1^K\mathbb{S}_{III}(u_i,v_i)}{\prod\limits_{i,j=1}^K
\left[u_i-\zeta_{\sigma(j)}\right]\left[v_i-\zeta_{\sigma(K+j)}\right]}\nonumber
\end{equation}
where $\sigma\;\in\;{\textsf{S}}_{2K}/{\textsf{S}}_{K}\times
{\textsf{S}}_{K}$ should be understood as permutations exchanging
elements between the two sets, $(i\delta/2,\ldots ,i\delta/2) $ and
$(-i\delta/2,\ldots ,-i\delta/2)$ (the dimension of each set is
$K$). Consider all such permutations that replace $K_1\leq K$
elements of the first set, $(i\delta/2,\ldots ,i\delta/2)$, by
$K_1$ elements of the second set, $(-i\delta/2,\ldots
,-i\delta/2)$. The number of such permutations is equal to
$\left[\frac{K!}{(K-K_1)!K_1!}\right]^2$. A precise location of
 new elements in the first and the second sets does not affect
the integral, so all $\left[\frac{K!}{(K-K_1)!K_1!}\right]^2$ give
the same contribution. Therefore,
\begin{equation}\label{NM72}
\J_K(\delta) =K!\;\sum\limits_{K_1=0}^K
\left[\frac{K!}{(K-K_1)!K_1!}\right]^2\;I^{K,K_1}_{\delta}
\end{equation}
where
\begin{equation}
I^{K,K_1}_{\delta} = \oint\oint
\prod\limits_1^K\frac{du_idv_i}{(2\pi i)^2} \frac{\triangle(\hat
u) \triangle(\hat v)\prod_1^KS_{III}(u_i,v_i)}{\prod\limits_{1}^K
\left[(u_l+ i\frac{\delta}{2})(v_l-
i\frac{\delta}{2})\right]^{K_1}\left[(u_l- i\frac{\delta}{2})(v_l+
i\frac{\delta}{2})\right]^{K-K_1}}\nonumber
\end{equation}
Now we rewrite the contour integral $I_{\delta}^{K,K_1}$ as a
determinant of a kernel divided by two Vandermonde determinants.
It follows from explicit expressions for limiting kernels
summarized in table 2 that $\mathbb{S}_{III}(\alpha,\beta)=0$ to the
leading order, if the variables $\alpha$ and $\beta$ have
imaginary parts of different signs. Then we obtain
\begin{equation}
I^{K,K_1}_{\delta}=\lim\limits_{\delta_i\rightarrow
\delta/2}\nonumber
\end{equation}
\begin{tiny}
\begin{equation}
\frac{\mbox{det} \left|
\begin{array}{cccccc}
  \mathbb{S}_{III}(-\frac{i\delta_1}{2},\frac{i\delta_1}{2}) & \ldots & \mathbb{S}_{III}(-\frac{i\delta_1}{2},\frac{i\delta_{K_1}}{2}) & 0 & \ldots & 0 \\
  \vdots &  &  &  &  &  \\
  \mathbb{S}_{III}(-\frac{i\delta_{K_1}}{2},\frac{i\delta_1}{2}) & \ldots & \mathbb{S}_{III}(-\frac{i\delta_{K_1}}{2},\frac{i\delta_{K_1}}{2}) & 0 & \ldots & 0 \\
  0 & \ldots & 0 & \mathbb{S}_{III}(\frac{i\delta_{K_1+1}}{2},-\frac{i\delta_{K_1+1}}{2}) & \ldots & \mathbb{S}_{III}(\frac{i\delta_{K_1+1}}{2},\frac{-i\delta_{K}}{2}) \\
  \vdots &  &  &  &  &  \\
0 & \ldots & 0 &
\mathbb{S}_{III}(\frac{i\delta_{K}}{2},-\frac{i\delta_{K_1+1}}{2})
& \ldots &
\mathbb{S}_{III}(\frac{i\delta_{K}}{2},\frac{-i\delta_{K}}{2})
\end{array}
\right|}{ \triangle \left(-\frac{i\delta_1}{2},\ldots
,-\frac{i\delta_{K_1}}{2},\frac{i\delta_{K_1+1}}{2},\ldots
,\frac{i\delta_{K}}{2}\right)\triangle\left(\frac{i\delta_1}{2},\ldots
,-\frac{i\delta_{K_1}}{2},-\frac{i\delta_{K_1+1}}{2},\ldots
,-\frac{i\delta_{K}}{2}\right)} \nonumber
\end{equation}
\end{tiny}
We insert the explicit expression for the kernel
$\mathbb{S}_{III}(\zeta-\eta)$ (see Table 2) to the above formula
and find that
\begin{equation}
I^{K,K_1}_{\delta}=\frac{
(-i)^KK!}{(\delta)^{2K_1(K-K_1)}}\nonumber
\end{equation}

\begin{equation}\label{IKDELTA}
\qquad\;\;\;\times\;\;\lim\limits_{\delta_i\rightarrow \delta}
\left[ \frac{
\mbox{det}\left(\frac{2}{\delta_i+\delta_j}\right)_{1\leq i,j\leq
K_1}}{\triangle^2\left(\frac{\delta_1}{2},\ldots
,\frac{\delta_{K_1}}{2}\right)}\;\;\frac{
\mbox{det}\left(\frac{2}{\delta_i+\delta_j}\right)_{K_1+1\leq
i,j\leq K}}{\triangle^2\left(\frac{\delta_{K_1+1}}{2},\ldots
,\frac{\delta_{K}}{2}\right)}\right]
\end{equation}
Exploiting once again Brezin and Hikami's representation in terms
of a contour integral we find
\begin{equation}
\lim\limits_{\delta_i\rightarrow \delta} \left[ \frac{
\mbox{det}\left(\frac{2}{\delta_i+\delta_j}\right)_{1\leq i,j\leq
M}}{\triangle^2\left(\frac{\delta_1}{2},\ldots
,\frac{\delta_{M}}{2}\right)}\right]=\mbox{det}\left(a_{mn}\right)_{0\leq
m,n\leq M-1}
\end{equation}
where
\begin{equation}
a_{mn}=\frac{1}{n!\;m!}\;\frac{\partial^n}{\partial u^n}
\frac{\partial^{m}}{\partial
v^m}\left[\frac{1}{u+v+\delta}\right]_{u=v=0}\nonumber
\end{equation}

\begin{equation}
\qquad=\frac{(-)^{m+n}}{\delta^{m+n}}\left(\frac{(m+n)!}{m!\;n!}\right)
\nonumber
\end{equation}
Now it is straightforward to compute
$\mbox{det}\left(a_{mn}\right)$. We obtain
\begin{equation}
\lim\limits_{\delta_i\rightarrow \delta} \left[ \frac{
\mbox{det}\left(\frac{2}{\delta_i+\delta_j}\right)_{1\leq i,j\leq
M}}{\triangle^2\left(\frac{\delta_1}{2},\ldots
,\frac{\delta_{M}}{2}\right)}\right]=\frac{1}{\delta^{M^2}}\nonumber
\end{equation}
This equation  together with the equation (\ref{IKDELTA}) yields
\begin{equation}
I_{\delta}^{K,K_1}=(-i)^K\; K!\;\delta^{-K^2}
\end{equation}
(i.e. $I_{\delta}^{K,K_1}$ does not depend on $K_1$). Taking into
account that
\begin{equation}
\sum\limits_{K_1=0}^K\left[\frac{K!}{(K-K_1)!K_1!}\right]=
\frac{(2K)!}{(K!)^2}
\end{equation}
 and equations (\ref{NM71}), (\ref{NM72}) we finally obtain our
result (\ref{NegativeMomentAsymptotic}) for the negative moments
of the characteristic polynomials.
\begin{flushleft}\end{flushleft}
\section{\textbf{Two-point Correlation Function}}
\begin{flushleft}\end{flushleft}
The present section serves for illustrating the utility of
the correlation functions constructed with help of
the characteristic polynomials.
We are going to demonstrate how one can derive
the resolvent two-point correlation
function from our asymptotic result for the average values of ratios of
the characteristic polynomials. We consider the case when
$\alpha(x)\equiv \frac{V'(x)}{2\rho(x)}=0$. Then  the correlation
function $\F_{II}^K(\hat x+\hat\zeta/N\rho(x),\hat x+
\hat\eta/N\rho(x))$ is universal in the Dyson scaling limit. As the
result, the
answer for the resolvent correlation function  will be universal
as well. as a by-product we provide a new proof of the universality of
the two-point correlation function of eigenvalue densities.

The resolvent two-point correlation function is defined as
\begin{equation}
S_2(E,E')=N^{-2}\left\langle\mbox{Tr}\left[\frac{1}{E-H}\right]
\mbox{Tr}\left[\frac{1}{E'-H}\right]\right\rangle_H
\end{equation}
where the (complex) energy $E$ has a positive imaginary part and the
 energy $E'$ has a negative imaginary part. To consider the
scaling limit we introduce   new coordinates
\begin{equation}
E=x+\eta_1/N\rho(x),\;\; \Im m\; \eta_1>0 \;
\end{equation}
\begin{equation}
E'=x+\eta_2/N\rho(x),\;\; \Im m\; \eta_2<0
\end{equation}
To connect the two-point
correlation function $S_2(x+\eta_1/N\rho(x),x+\eta_2/N\rho(x))$
with the correlation function $\F_{II}^{K=2}(\hat
x+\hat\zeta/N\rho(x),\hat x+ \hat\eta/N\rho(x))$ investigated
above we exploit the following identity
\begin{equation}
\mbox{Tr}\left[\frac{1}{x+\eta/N\rho(x)-H}\right]=\left[N\rho(x)\right]\;
\frac{\partial_{\eta}\Z_{N}[x_{\eta},H]}{\Z_{N}[x_{\eta},H]}
\end{equation}
where $x_{\eta}\equiv x+\eta/N\rho(x)$ . We have
\begin{equation}
S_2(x+\eta_1/N\rho(x),x+\eta_2/N\rho(x))\nonumber
\end{equation}
\begin{equation}\qquad\qquad=\left[\rho(x)\right]^2\left[\partial_{\eta_1,\eta_2}^2
\left\langle
\frac{\mbox{det}\left(x_{\eta_1}-H\right)\mbox{det}\left(x_{\eta_2}-H\right)}
{\mbox{det}\left(x_{\zeta_1}-H\right)\mbox{det}\left(x_{\zeta_2}-H\right)}
\right\rangle\right]_{\zeta_1=\eta_1,\zeta_2=\eta_2} \nonumber
\end{equation}

\begin{equation}
\qquad\qquad
=\left[\rho(x)\right]^2\left[\partial_{\eta_1,\eta_2}^2
\F_{II}^{K=2}(\hat x+\hat\zeta/N\rho(x),\hat x+ \hat\eta/N\rho(x))
\right]_{\zeta_1=\eta_1,\zeta_2=\eta_2} \nonumber
\end{equation}
 Equation (\ref{DysonLimitFII}) yields
\begin{equation}
\F_{II}^{K=2}(\hat \hat x+\hat\zeta/N\rho(x),\hat x+
\hat\eta/N\rho(x))=\frac{e^{i\pi(\zeta_1-\zeta_2)}}{\zeta_1-\zeta_2}
\nonumber
\end{equation}

\begin{equation}
\times\left[e^{i\pi(\eta_1-\eta_2)}\frac{(\eta_1-\zeta_1)(\eta_2-\zeta_2
)}{\eta_1-\eta_2}-e^{-i\pi(\eta_1-\eta_2)}\frac{(\eta_1-\zeta_2)(\eta_2-\zeta_1
)}{\eta_1-\eta_2}\right] \nonumber
\end{equation}
The next step is to compute derivatives. In particular, we find
\begin{equation}
\partial^2_{\eta_1,\eta_2}\left[
e^{i\pi(\eta_1-\eta_2)}\frac{(\eta_1-\zeta_1)(\eta_2-\zeta_2
)}{\eta_1-\eta_2}\right]_{\zeta_1=\eta_1,\zeta_2=\eta_2}\nonumber
\end{equation}

\begin{equation}
\qquad\qquad\qquad\qquad\qquad\qquad\qquad=\frac{e^{i\pi(\eta_1-\eta_2)}}{\eta_1-\eta_2}
\end{equation}
and
\begin{equation}
\partial^2_{\eta_1,\eta_2}\left[
e^{-i\pi(\eta_1-\eta_2)}\frac{(\eta_1-\zeta_2)(\eta_2-\zeta_1
)}{\eta_1-\eta_2}\right]_{\zeta_1=\eta_1,\zeta_2=\eta_2}\nonumber
\end{equation}

\begin{equation}
\qquad\qquad\qquad\qquad\qquad\qquad\qquad=-\pi^2(\eta_1-\eta_2)e^{-i\pi(\eta_1-\eta_2)}+
\frac{e^{-i\pi(\eta_1-\eta_2)}}{\eta_1-\eta_2}
\end{equation}
which gives the well-known result (see \cite{mehta,Haake})  for the resolvent
two-point correlation function:

\begin{equation}
S_2(x+\eta_1/N\rho(x),x+\eta_2/N\rho(x))\nonumber
\end{equation}

\begin{equation}
\qquad\qquad\qquad=\left[\pi\rho(x)\right]^2
\left[1-2i\;\frac{\sin\pi(\eta_2-\eta_1)\;e^{-i\pi(\eta_2-\eta_1)}
}{\left[\pi(\eta_2-\eta_1)\right]^2}\right]
\end{equation}
\begin{flushleft}\end{flushleft}
\section{\textbf{Summary and Discussions}}
\begin{flushleft}\end{flushleft}
In this paper we prove three basic statements: 1) correlation
functions of characteristic polynomials are governed by two-point
kernels, 2) the kernels are "integrable" in the sense of the
definition by Its,
Izergin, Korepin and Slavnov
\cite{itsizergin1}-\cite{itsizergin3}, 3) the kernels are
constructed from monic orthogonal polynomials and their Cauchy
transforms. As a consequence, it becomes quite natural to exploit a
relation to the Riemann-Hilbert problem for orthogonal
polynomials  proposed by Fokas, Its and Kitaev
\cite{fokas1,fokas2}.

It is known that the simplest correlation functions, i.e. the
moments of the characteristic polynomials, can be described in
terms of non-linear differential equations (see works of Forrester
and White \cite{forrester0,forrester1}, and also the paper by
Kanzieper \cite{kanzieper}, and by Splittorff and Verbaarschot
\cite{Verb}). As for more complicated correlation functions a
description in terms of differential equations is unknown. While
in the present paper we focus on  asymptotic questions, the
"integrability" of kernels  suggests that such description should
be possible.

In this paper the discussion is restricted to  the ensemble of
unitary invariant Hermitian matrices. However, the case of
compact group ensembles (circular ensembles) can be approached in
the same way. The circular ensembles are even simpler as the
representation theory of compact groups (inapplicable for
Hermitian random matrices) can be exploited there.  Indeed, the
method of "dual pairs" (Nonenmacher and Zirnbauer
\cite{ZN,zirnbauer1,zirnbauer2}) enables one to find exact formulas
for correlation functions of characteristic polynomials for the
circular ensembles. However, this method is  based on an
interpretation of characteristic polynomials as characters of
"spinor" group representations. As a result, its applicability is
restricted to group ensembles. The method proposed in this paper
is based neither on the representation theory, nor on specific
features of Hermitian matrices. The only fact which is important
is that the ensemble under considerations is of $\beta=2$ symmetry
class. For this reason it is more general and can be applied both
to Hermitian random matrices and to the group ensembles.

Correlation functions of characteristic polynomials of random
matrices for the unitary circular ensemble  and the ensemble of
Hermitian matrices are sometimes represented  as Toeplitz / Hankel
determinants, correspondingly.  Results for Toeplitz determinants
with rational generating functions can be found in Basor and
Forrester \cite{basor1}. However, Hankel determinants are less
well studied. The correlation functions investigated in the
present paper can be understood as Hankel determinants with
rational generating functions. Thus, asymptotic and exact
statements for Hankel determinants equivalent to our results
should be possible to make.

We hope that the method proposed in this paper can be modified as
to provide an access to other symmetry classes of invariant
ensembles $\beta=1,4$ (for the gaussian case some attempts were
undertaken recently in \cite{brezin3}).
 Another important goal is to apply them to ensembles of
non-Hermitian random matrices \cite{FK, F11}, which are certain
deformations of the invariant class.

The most challenging problem  is to investigate the conditions of
universality of the discussed correlation functions for
non-invariant non-gaussian ensembles with independent, identically
distributed (i.i.d.) entries. So far the progress was rather
limited and restricted to the specific choice of the probability
measure (see Johansson \cite{Johuni}). At the same time,
non-rigorous heuristic methods hint to a kind of universality
covering also the so-called ensembles of sparse random matrices,
see Mirlin, Fyodorov \cite{MF} and Fyodorov, Sommers \cite{FScur}.

We leave a detailed investigation of these issues for future research.
\begin{flushleft}\end{flushleft}
\section*{\textbf{Acknowledgements}}
\begin{flushleft}\end{flushleft}
We would like to thank A. Its, A. Kamenev, J. P. Keating, B. A.
Khoruzhenko and L. A. Pastur for their interest, valuable comments
and discussions on various stages of the work. We are grateful to
I. Krasikov for his remark on a relation to the Lagrange
interpolation formula. This research was supported by EPSRC grant
GR/13838/01 "Random matrices close to unitarity or Hermitian".
\begin{flushleft}\end{flushleft}
\begin{flushleft}\end{flushleft}
\begin{flushleft}\end{flushleft}
\appendix
\section{\textbf{}}
\begin{flushleft}\end{flushleft}
Let us consider the sum
\begin{equation}
S=\sum\limits_{\sigma,\;\pi\;\in\;
{\textsf{S}}_{n+m}}(-)^{\nu_{\sigma+\pi}}[f_{\sigma(1)}(\varphi_1)
\ldots f_{\sigma(m)}(\varphi_m)\nonumber
\end{equation}


\begin{equation}\label{sum1}
 \qquad\qquad\qquad\;\;\;\times\;\qquad\;  g_{\pi(1)}(\psi_1)
\ldots  g_{\pi(m)}(\psi_m)
\end{equation}

\begin{equation}
\qquad\qquad\qquad\;\;\;\times\;\qquad\;
\delta_{\sigma(m+1)\pi(m+1)}\ldots
\delta_{\sigma(m+n)\pi(m+n)}]\nonumber
\end{equation}
In order to reduce this expression to a determinant form (to a
determinant of size $m\times m$) we proceed as follows. We fix the
set of numbers $[k_1,\ldots ,k_m]$ satisfying the following
condition
\begin{equation}
n+m\geq k_1>k_2>\ldots k_m\geq 1
\end{equation}
Let us denote by $\tilde\sigma(1),\ldots ,\tilde\sigma(m)$ ( and
by $\tilde\pi(1),\ldots ,\tilde\pi(m)$) the permutations under
which the numbers $1,2,\ldots ,m$ end up inside the set
$[k_1,k_2,\ldots ,k_m]$. Once such notations are introduced, the
sum $S$ (equation (\ref{sum1})) can be rewritten as
\begin{equation}
S=\sum\limits_{k_1>\ldots >k_m\geq 1}^{n+m}\;\;\;
\sum\limits_{\tilde\sigma,\;\tilde\pi\;\in\;
{\textsf{S}}_{n+m}}(-)^{\nu_{\tilde\sigma+\tilde\pi}}[f_{\tilde\sigma(1)}(\varphi_1)
\ldots f_{\tilde\sigma(m)}(\varphi_m)\nonumber
\end{equation}

\begin{equation}
\qquad\qquad\qquad\qquad\qquad\qquad\;\;\;\times\; \qquad
g_{\tilde\pi(1)}(\psi_1) \ldots g_{\tilde\pi(m)}(\psi_m)
\end{equation}

\begin{equation}
\qquad\qquad\qquad\qquad\qquad\qquad\;\;\; \times\;
\qquad\delta_{\tilde\sigma(m+1)\tilde\pi(m+1)}\ldots
\delta_{\tilde\sigma(m+n)\tilde\pi(m+n)}]\nonumber
\end{equation}
Under permutations $\tilde\sigma ,\tilde\pi$ the numbers
$m+1,\ldots ,m+n$ remain outside of the set $[k_1,\ldots ,k_m]$.
The sets $[1,\ldots ,m]$, $[m+1,\ldots ,m+n]$ do not mix, and this
leads to the following expression for $S$:
\begin{eqnarray}
S=\sum\limits_{k_1>\ldots >k_m\geq 1}^{n+m}\;\;\;
\left[\sum\limits_{\tilde\sigma\;\in\;
{\textsf{S}}_{m}}(-)^{\nu_{\tilde\sigma}}f_{k_{\tilde\sigma(1)}}(\varphi_1)
\ldots f_{k_{\tilde\sigma(m)}}(\varphi_m)\right]\\
\times\; \left[\sum\limits_{\tilde\pi\in\;{\textsf{S}}_{m}}
(-)^{\nu_{\tilde\pi}}g_{k_{\tilde\pi(1)}}(\psi_1) \ldots
g_{k_{\tilde\pi(m)}}(\psi_m)\right]\nonumber\\
\times\;\left[\sum\limits_{\tilde\sigma,\;\tilde\pi\;\in\;
{\textsf{S}}_{n}}(-)^{\nu_{\tilde\sigma}+\nu_{\tilde\pi}}
\delta_{\tilde\sigma(m+1)\tilde\pi(m+1)}\ldots
\delta_{\tilde\sigma(m+n)\tilde\pi(m+n)}\right]\nonumber
\end{eqnarray}
From the above representation we immediately conclude that
\begin{equation}\label{detdet}
S=n!\;\sum\limits_{k_1>\ldots >k_m\geq 1}^{n+m}\;
\mbox{det}\left[f_{k_i}(\varphi_j)\right]\;
\mbox{det}\left[g_{k_i}(\psi_j)\right]
\end{equation}
where the indices $i,j$ take values from $1$ to $m$. Now the sum
$S$ can be rewritten as a determinant $m\times m$, as a
consequence of   the formula
\begin{equation}\label{Sdeterminant}
\sum\limits_{k_1>\ldots >k_m\geq 1}^{n+m}\;
\mbox{det}\left[f_{k_i}(\varphi_j)\right]\;
\mbox{det}\left[g_{k_i}(\psi_j)\right]
=\mbox{det}\left[\sum\limits_{\lambda=0}^{n+m-1}f_{\lambda}(\varphi_i)
g_{\lambda}(\psi_j)\right]_{1\leq i,j\leq m}
\end{equation}
(This fact is a generalization of the theorem on determinants.
This theorem ( which can be found in the book of Hua \cite{hua})
states that
\begin{equation}
\mbox{det}\left[\sum\limits_{\lambda=0}^{+\infty}A_{\lambda,j}t_i^{\lambda}
\right]_{1\leq i,j\leq m} =\sum\limits_{\lambda_1 > \ldots >
\lambda_m\geq
0}^{+\infty}\mbox{det}\left[A_{\lambda_j,i}\right]_{1\leq i,j\leq
m}\det\left[t_i^{\lambda_j}\right]_{1\leq i,j\leq m}
\end{equation}
A nice proof of the formula (\ref{Sdeterminant}) can be found, for
example, in the paper by Balantekin and Cassak \cite{balantekin}).

\end{document}